\documentclass[aps,prx,longbibliography,twocolumn,amsmath,amssymb,floatfix,eqsecnum,superscriptaddress]{revtex4-1}

\usepackage{amsmath}
\usepackage{amssymb}
\usepackage{amsthm}
\usepackage[pdftex]{color} 
\usepackage{graphicx}
\usepackage{dcolumn} 
\usepackage{bm} 
\usepackage{ulem}

\usepackage[driverfallback=dvipdfm]{hyperref} 

\usepackage{longtable}
\usepackage{ulem}   
\newcommand{\ket}[1]{{|{#1}\rangle}}
\newcommand{\bra}[1]{{\langle{#1}|}}

\renewcommand{\vec}[1]{\bm{#1}}

\newcommand{\be} {\begin{equation}}
\newcommand{\ee} {\end{equation}}

\begin{document}

\title{Pair-Density-Wave Order and Paired Fractional Quantum Hall Fluids}
      
\author{Luiz H. Santos}
\thanks{
These authors contributed equally to the work.}
\affiliation
{
Department of Physics and Institute for Condensed Matter Theory,
University of Illinois at Urbana-Champaign, 1110 West Green Street, Urbana, Illinois, 61801-3080, USA
}
\affiliation
{
Department  of  Physics,  Emory  University,  400 Dowman Drive, Atlanta,  GA  30322,  USA
}

\author{Yuxuan Wang}
\thanks{
These authors contributed equally to the work.}
\affiliation
{
Department of Physics and Institute for Condensed Matter Theory,
University of Illinois at Urbana-Champaign, 1110 West Green Street, Urbana, Illinois, 61801-3080, USA
}
\affiliation
{
Department of Physics, University of Florida, 2001 Museum Rd, Gainesville, FL 32611
}

\author{Eduardo Fradkin}
\affiliation
{
Department of Physics and Institute for Condensed Matter Theory,
University of Illinois at Urbana-Champaign, 1110 West Green Street, Urbana, Illinois, 61801-3080, USA
}

\begin{abstract}

The properties of the isotropic incompressible {$\nu=5/2$} fractional quantum Hall (FQH) state
are described by a paired state of composite fermions in zero (effective) magnetic
field, with a uniform $p_x+ip_y$ pairing order parameter, which is a non-Abelian topological phase
with chiral Majorana and charge modes at the boundary.
Recent experiments suggest the existence of a proximate nematic phase at $\nu=5/2$.  This finding motivates us to consider an inhomogeneous 
paired state --- a $p_x+ip_y$ pair-density-wave (PDW) --- whose melting could be the origin of the observed liquid-crystalline phases.
This state can viewed as an array of domain and anti-domain walls of the $p_x+i p_y$ order parameter.
We show that the nodes of the PDW order parameter, {the location of the  domain walls (and anti-domain walls)} where the order parameter changes sign, 
support a pair of symmetry-protected counter-propagating Majorana modes. The coupling behavior of the domain wall Majorana modes crucially depends on the interplay
of the Fermi energy $E_{F}$ and the PDW pairing energy $E_{\textrm{pdw}}$.
The analysis of this interplay yields a rich set of topological states:
(1) In the weak coupling regime ($E_F > E_{\rm pdw}$), the hybridization of domain walls leads to a Majorana Fermi surface (MFS), which is protected by inversion and particle-hole symmetries. 
(2) As the MFS shrinks towards degenerate Dirac points, lattice effects render it unstable towards an 
Abelian striped phase with two co-propagating Majorana modes at the boundary. 
(3) An uniform component of the order parameter, which breaks inversion symmetry, gaps the MFS and causes the system to enter a non-Abelian FQH state supporting a chiral Majorana edge state. 
(4) In the strong coupling regime, $E_F < E_{\rm pdw}$, the bulk fermionic spectrum becomes gapped; 
this is a trivial phase with no chiral Majorana edge states, which is in the universality class of an Abelian Halperin paired state.
The pair-density-wave order state in paired FQH system provides a fertile setting to study Abelian and non-Abelian FQH phases --- as well as transitions thereof --- tuned by the strength of the paired liquid crystalline order.
\end{abstract}

\date{\today}

\maketitle
 \tableofcontents

\section{Introduction}
\label{sec: introduction}

Fractional Quantum Hall (FQH) states are the quintessential example 
of topological electronic systems.
While the majority of the FQH plateaus are observed near filling fractions $\nu = p/q$ with odd denominators \cite{Prange-book}, 
even-denominator FQH states \cite{willett87,pan99} 
provide a fertile arena to study exotic non-Abelian statistics \cite{mooreread91,nayakwilczek96}, 
as well as the interplay between symmetry breaking and topological orders.

 In addition to FQH states, a host of symmetry breaking states have also been observed in two-dimensional electron gases (2DEGs) in magnetic field in various Landau levels (LL). These states, generally known as electronic liquid crystal phases, \cite{KFE,FK} break spatial symmetries to various degrees. Examples of such states are crystals (Wigner crystals \cite{Prange-book} and bubble phases \cite{Fogler-1996}), stripe phases \cite{moessner-1996,Koulakov-1996,FK}, and electronic nematic states \cite{FK,fradkin-2000}. While crystal phases break translation and rotational invariance (down to the point group  symmetry of the underlying lattice), stripe (or smectic) phases break translation invariance along one direction (and concomitantly rotation symmetry), nematic phases only break rotational invariance  and are spatially uniform
  \cite{FKLEM}.  Most of the stripe and nematic phases that have so far been seen in experiment are compressible, and do not exhibit the (integer or fractional) quantum Hall effect, although they occur in close proximity to such incompressible states. Compressible  nematic phases exhibit strong transport anisotropies, which is how they are detected experimentally. In addition, stripe phases also exhibit strong pinning and non-linear transport at low bias. Compressible electronic nematic order was first observed
at filling fractions in $N \geq 2$ LL such as $\nu = 9/2, 11/2$, etc.~\cite{lilly99-1,du99,pan2014} 
Evidence for transition from a stripe to a nematic order in the $N=2$ LL in a compressible regime has also been seen quite recently.~\cite{qian2017}

On the other hand, in the $N=1$ LL, FQH states observed \cite{willett87} at $\nu=5/2$ are presumably paired states of the Moore-Read type \cite{mooreread91}. In addition, other paired FQH states have been proposed to explain the plateau observed at $\nu=5/2$ \cite{Levin-2007,SSLee-2007,
Wang-2018,Mross-2018,lian-wang2018}. Remarkably,  experimental results in the $N=1$ LL also show the existence of states with nematic order, originally in samples  where rotation symmetry is explicitly broken by an in-plane 
magnetic field \cite{pan99-2,lilly99-2,friess2014,shi2015,xia2010,liu2013}. 
More recently, a spontaneously formed nematic phase has been reported 
in GaAs/AlGaAs samples under hydrostatic pressure \cite{samkharadze2016}. (See also Refs.~\cite{schreiber2017} and~\cite{schreiber2018}.)
The mechanism behind this spontaneous nematicity remains an open problem, and has been speculated that it could 
 be due to a Pomeranchuk instability of the composite fermions, as indicated by recent a numerical calculation.~\cite{lee2018}
In all of these experiments the nematic phase is compressible, arises after the gap or the $5/2$ FQH state has vanished, and it does not have a quantum Hall plateau.
Magnetoresistance measurements show that  the isotropic $5/2$ FQH state collapses at a  hydrostatic pressure $P_c \approx 7.8$ kbar. This is followed by onset of a compressible nematic state detected as a strong and temperature-dependent longitudinal transport anisotropy at higher  pressures. This nematic phase persists up to a critical value of $10$ kbar, where the 2DEG appears to become a Fermi liquid.

In addition, experiments also discovered in the $N=1$ LL   a large nematic susceptibility (with a strong temperature dependence)  in the FQH state with $\nu=7/3$ \cite{xia2011}. 
This experimental finding suggests that, at least in the $N=1$ Landau level, nematic and/or stripe order may also occur in {proximity and/or coexistence} with a FQH topological state. 

The experimental observation of (presumably) paired FQH states in close proximity to nematic, and possibly stripe, phases suggests that all these phases may have a common physical origin, and that these orders may be actually intertwined rather than simply competing with each other. This scenario is strongly reminiscent of the current situation in cuprate superconductors, and other strongly correlated oxides, where superconducting orders are intertwined, rather than competing, with stripe or nematic phases \cite{Berg-2009,Fradkin2015}. The prototype of an intertwined superconducting state is a superconducting state known as the pair-density-wave (PDW) \cite{Berg-2007}. The PDW is a paired state that breaks spontaneously translation invariance. Its order parameter is closely related to that of  the Larkin-Ovchinnikov state (although occurring in the absence of  a Zeeman coupling to an external magnetic field).

 A system of electrons in a half-filled Landau level, such as the $N=1$ LL in the case of the $5/2$ FQH state, is equivalent to a system of composite fermions \cite{Wilczek-1982,Jain1989} coupled to a Chern-Simons gauge field, in which  two flux quanta have been attached to each electron \cite{Lopez1991}. The  composite fermions  are coupled to both the external magnetic field and to the dynamical Chern-Simons gauge field.  In a half-filled Landau level, the composite fermions experience, on average, an effective zero magnetic field. The resulting (mean field state) forms a Fermi surface (FS) of composite fermions \cite{HLR}. In this representation, the topological  incompressible isotropic FQH at $\nu = 5/2$ arises from a pairing instability of the composite fermion Fermi surface, resulting in a chiral paired state. In other terms, the paired FQH state can be viewed as a superconductor with $p_x+ip_y$ pairing coupled to a dynamical Chern-Simons gauge field at  level 2.

The aim of this paper is to construct an intertwined orders scenario for a 2DEG proximate to a paired Moore-Read state \cite{mooreread91} near the $\nu=5/2$ filling fraction. 
The state that we will propose is a stripe state that locally has a $p_x+ip_y$ form while, at the same time, 
breaking unidirectional translation invariance. We will call the resulting intertwined state a $p_x+ip_y$ paired density wave state (instead of the $d$-wave local pairing of the PDW state of the cuprate high $T_c$ superconductors.)
Such a state may also occur as an inhomogeneous version of  a topological $p_x+ip_y$ superconductor as well. Here we will not consider  other typed paired states  \cite{Levin-2007,SSLee-2007,Mross-2018}. Non-translational invariant states  with  local $p_x \pm ip_y$ pairing were considered recently as a possible way to restore the Landau level particle-hole symmetry (broken by both the pfaffian and the anti-pfaffian paired states) \cite{Wan-2016}, as well as a  paired and particle-hole symmetric state driven by disorder  \cite{Wang-2018}. These states have different physical properties than those of the $p_x+ip_y$ PDW state we present in this paper.

In this work we  formulate a theory of a $p_x+ip_y$ PDW state, which is an interesting superconducting state in its own right, and later examine the resulting  FQH state by considering the effects of coupling this PDW state to the dynamical Chern-Simons gauge field. 
The resulting state has the remarkable property of having neutral fermionic excitations that are either gapless or gapped (with nontrivial band topology). 
{In the gapless case, the neutral (Majorana) fermions form Fermi surfaces and, hence, has a finite bulk thermal conductivity.}
At the same time, {this state is  incompressible in the charge channel,  has a plateau with a precisely defined  Hall conductivity, and protected chiral charge edge states.} 

The $p_x+ip_y$ PDW FQH state  can be viewed as an array of stripes of Moore-Read states in which the $p_x+ip_y$ pair field changes sign from one stripe to the next, 
in close analogy with  the  (time-reversal invariant)  PDW superconductor discussed in the context of high temperature superconductors \cite{Berg-2009}. 
This unidirectional state breaks translation invariance along one direction and also breaks rotations by $90^\circ$. 
Since locally it is equivalent to a Moore-Read state, this state also breaks the particle-hole symmetry of the Landau level as well. 
The $p_x+ip_y$ PDW FQH state can arise either by spontaneous symmetry breaking of translation (and rotation) symmetry, 
or by the explicit breaking of rotation symmetry by  a tilted-magnetic field, or by in-plane strain, 
as in the very recent experiments by Hossain and coworkers \cite{Hossain-2018}. 
The $p_x+ip_y$ PDW FQH breaks time-reversal symmetry as much as the uniform $p_x+ip_y$ paired state does. 
In contrast,  the particle-hole symmetric striped Pfaffian state proposed by 
Wan and Yang \cite{Wan-2016} consists of an array of alternating $p_x+ip_y$ and $p_x-ip_y$ stripes.
While both states break translation (and rotation) symmetry, the $p_x+ip_y$ PDW FQH state breaks the Landau level particle-hole symmetry 
whereas the Wan-Yang state does not. Thus, the $p_x+ip_y$ PDW breaks time reversal invariance explicitly, 
whereas in the Wan-Yang state time reversal is equivalent to a translation by half the period of the state. 
These differences lead to profound differences in their spectra and physical properties. 

{How is the $p_x+ip_y$ PDW state related to the observed phenomena in the $N=1$ Landau level? The physical picture that we propose is that this competitor of the uniform paired state may be responsible for the observed complexity of the phase diagram in these 2DEGs. To be more specific, upon quantum and/or thermal melting, this state may give rise to a sequence of phase transitions, much in the same way as envisioned in Ref.~\cite{FK}. Thus, upon melting the stripe order, a nematic state may ensue which may coexist with the FQH state, similar to what was found in the experiments of Xia and coworkers \cite{xia2011} in the $\nu=7/3$ plateau, or in a compressible regime, as in the pressure-driven experiments of Samkharadze and coworkers \cite{samkharadze2016} in the regime with $\nu=5/2$ filling fraction. From this perspective, the nematic order would not be a ``primitive'' order, but a ``vestigial'' one. An example of such a phase diagram was studied in the context of the (proposed) PDW superconducting state for the cuprates \cite{Berg-2009b}. In this sense we can think of the compressible nematic state observed in Ref. \cite{samkharadze2016} as a form of vestigial order of a putative $p_x+ip_y$ PDW paired state. We note that theories of Laughlin FQH states that coexist with nematic order have also been discussed in the literature~\cite{Joynt-1996,Balents-1996,Mulligan-2010, Mulligan-2011, Maciejko-2013,You-2014,Nguyen-2018}.}

{There have been a great deal of work on the role of particle-hole symmetry vis-a-vis the paired FQH states. The Moore-Read Pfaffian state \cite{mooreread91} (and its anti-Pfaffian cousin \cite{Levin-2007,SSLee-2007}) is not particle-hole symmetric. The  $p_x+ip_y$ PDW paired state presented here is based on the Moore-Read state and, as such, is not particle-hole symmetric either. Several particle-hole symmetric states have been proposed. One such state is the PH-Pfaffian state \cite{Son-2015}. Given the strong numerical evidence for both Pfaffian and anti-Pffafian, inhomogeneous states that restore particle-hole symmetry on large scales, while breaking it at the local level, have been proposed. Two of these proposals advocate disorder as the driving mechanism \cite{lian-wang2018,Wang-2018}, while another one \cite{Wan-2016} proposed a stripe-type state, with alternating Pfaffian and anti-Paffian stripes. One feature of both the disorder-driven states and the PH symmetric striped state is a non-quantized thermal Hall conductivity coexisting with a quantized Hall conductivity. This is a feature also present in the  $p_x+ip_y$ PDW paired state but for a different reason: it harbors bulk (anisotropic) Fermi surfaces of Bogoliubov (Majorana) quasiparticles while supporting a chiral charge edge modes as in any FQH state.}

{Regarding the fluctuations of the PDW phase, because the PDW order parameter couples to fermions bilinearly, the fluctuations of the amplitude of the PDW order will mediate interaction effects of the Majorana modes. If such effects are weak, we expect the resulting interacting state to be similar to a Fermi liquid, only being electrically neutral, albeit with possible changes
in the Majorana bare velocities due to coupling with bosonic modes~\cite{sitte2009}. Thus the Majorana FS should remain robust if the amplitude fluctuations are sufficiently weak.}

{The  $p_x+ip_y$ PDW paired state is a two-dimensional state with spontaneously broken translation symmetry. As such, it is a fragile state to the effects of thermal fluctuations and disorder. In a strictly continuum system, dislocations of the broken translation symmetry cost a finite amount of energy. This causes this state (a smectic) to melt at any finite temperature resulting in a state with broken orientational invariance, a nematic \cite{Toner-1980}. This feature is also present in PDW superconductors \cite{Barci-2011}. However, coupling to an underlying lattice (however weakly, and provided the stripe order is incommensurate) makes the energy of a dislocation logarithmically divergent leading to a Kosterlitz-Thouless transition where the translation symmetry is fully restored. However, in this scenario, the orientational order is more robust since the associated symmetry is the point group symmetry of the lattice, a square lattice in the case of GaAs 2DEGs.}

{On the other hand, disorder has much more serious effects on states that break translation invariance, such as the  $p_x+ip_y$ PDW paired state. Strictly speaking, in two dimensions there is no true long range order of this type for any amount of disorder \cite{Imry-1975}, although the length scale over which disorder effects become prevalent may be exponentially long in a very clean system. In addition, disorder also couples to the non-chiral Majorana states bound to the domain walls we used to construct the $p_x+ip_y$ PDW paired state, by breaking  inversion symmetry locally. This breaking results in a  local, random, Majorana mass term for the states of each wall. This type of disorder does not lead to an Anderson insulator. Instead, for a single isolated wall, this system is equivalent to a critical transverse-field Ising model with quenched disorder \cite{mccoy-wu-1968,shankar-murthy-1987,fisher-1992}, which is known to be described by an infinite disorder fixed point \cite{fisher-1995}, in which the system remains critical. While this is true for an isolated wall, the behavior of a collection of such walls, coupled by tunneling of the Majorana fermions, has not been investigated, and remains an open problem. It is quite possible that the bulk 2D state may still be a thermal metal of sorts.}

This work is organized as follows. In Section \ref{sec:p_x+ip_y-PDW} we setup the proposed $p_x+ip_y$ PDW state and present a summary of the main results both as a possible superconducting state and as an inhomogeneous paired FQH state. In Section \ref{sec:Properties-Fermionic} we present a theory of the $p_x+ip_y$ paired state. Here we present the solution of the Bogoliubov-de Gennes (BdG) equations for this PDW state and discuss in detail the properties of its fermionic spectrum. 
In Section \ref{sec: Coexistence of stripe pairing order and uniform pairing order} 
we study the coexistence of the PDW order and the uniform pairing order.
In Section \ref{sec:PDW-FQH} we use this construction to infer the properties of the $p_x+ip_y$ PDW FQH state.  Section \ref{sec:conclusions} is devoted to the experimental implications of this PDW state and to conclusions. 
Theoretical details are presented in Appendices \ref{app:nodes}, \ref{app:stability}, \ref{app:tight-binding} and \ref{app:e-m}.

\begin{figure*}[t]
\includegraphics[width=2.05\columnwidth]{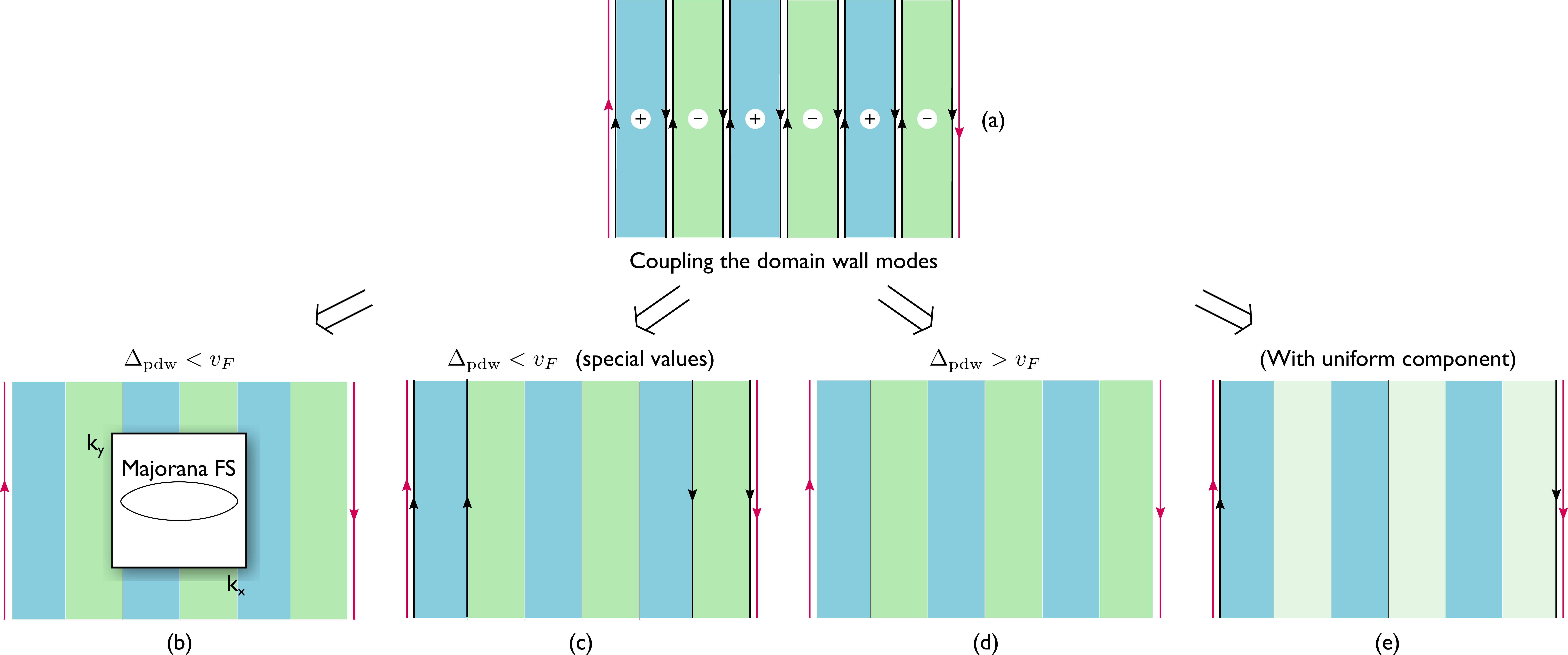}
\caption{{Illustration of the different quantum Hall states from the stripe pairing order. The blue and green strips denotes regions with positive and negative local pairing order parameter. The black arrows denote chiral Majorana modes while the red arrows denote chiral bosonic charge modes.
Panel (a): Gapless modes on domain walls of the pair density wave (PDW) and on physical edges in the limit of vanishingly small localization length and negligible couplings between these modes.  Panel (b): For PDW order parameter $\Delta_{\rm pdw}<v_F$, in general the domain wall modes form a Majorana FS, while there exist an energy gap for charge excitations. Panel (c): For particular values (see Sec.~\ref{subsubsec: c=2}) of PDW order parameter $\Delta_{\rm pdw}<v_F$, the Majorana FS shrinks to zero size and the fermionic sector gets gapped. In our model this phase has $\mathcal{C}=2$, and has Abelian topological order. Panel (d): For $\Delta_{\rm pdw}>v_F$, the fermionic sector become trivially gapped, and the resulting quantum Hall state is an Abelian one. 
 Panel (e): The neutral FS at $\Delta_{\rm pdw}<v_F$ becomes gapped with a uniform $p_x+ip_y$ pairing component with a nontrivial topology. The resulting quantum Hall state has non-Abelian topological order just as the Pfaffian state.}}
\label{fig: phases}
\end{figure*}

\section{The $p_x+ip_y$ pair density Wave: Setup and results}
\label{sec:p_x+ip_y-PDW}

In this section we present a summary on   
the $p_x+ip_y$ PDW state.
The pairing order parameter of the uniform $p_x+ip_y$ state has the form $\Delta(\vec{p}) = \Delta(p_x + ip_y)$ (with $\Delta = \textrm{constant}$). Its effective BdG Hamiltonian is
$
H = \sum_{\vec{p}}\,(\frac{\vec{p}^2}{2m} - \mu)\,\psi^{\dagger}_{\vec{p}}\,\psi_{\vec{p}}
+
\Delta(\vec{p})\,\psi_{-\vec{p}}\,\psi_{\vec{p}} + \textrm{H.c.}
$,
where $m$ is the composite fermion effective mass and $\mu$ is the chemical potential.~\cite{readgreen2000} 
In the ``weak-pairing phase'' of Ref.~\cite{readgreen2000}, where $\mu > 0$, 
this system is  a chiral topological superconductor where 
all bulk fermionic excitations are gapped and there is a chiral Majorana edge state propagating along 
the boundary separating the topological $p$-wave state and the vacuum.

The $p_x+ip_y$ PDW state that we propose here is a version of this state with a spatially modulated order parameter of the form
$\Delta\sim \Delta_{\textrm{pdw}}f(\vec Q\cdot \vec r)$, where $f$ is a periodic function with period $\lambda = 2\pi/Q$,
such that the nodes of $f$ correspond to domain walls (DWs) and anti-domain walls (ADWs), where the order parameter 
is suppressed, thus allowing for the existence of low energy modes localized on these nodes. Here, for simplicity, we consider only unidirectional order.

In the language of superconductors, our theory is analogous to the  PDW state conjectured for the cuprates whose order parameter has  wave vector $\vec{Q} = (Q,0)$ and that locally has $d$-wave superconducting (SC) order parameter \cite{Berg-2007,Agterberg-2008,Himeda-2002,Raczkowski-2007,Wang-2014,Lee-2014}. The main difference is that the PDW state that we consider here has, instead, local $p_x+ip_y$ pairing order. Although at the level of the Landau-Ginzburg theory the $d$-PDW and the $p_x+ip_y$-PDW are virtually identical, their fermionic spectra are drastically different as are their topological properties. 

Before moving forward with our analysis of this problem,
we stress important differences between the low energy fermion states we shall encounter in this work, which are
associated with the spatial modulation of the PDW order parameter, and those discussed by Read and Green.~\cite{readgreen2000}
As discussed in Ref.~\cite{readgreen2000}, the edge state of $p_x+ip_y$ paired state is a chiral Majorana fermion theory. The existence of this
chiral branch is of topological origin, since the edge represents a Chern number changing transition from $\mathcal{C} = 1$ (in the bulk of the 
paired state) to $\mathcal{C}=0$ (in vacuum). 
This change in the Chern number is also tied to the change in the sign of the chemical potential
in the BdG Hamiltonian, for the region with $\mu > 0$ is topological $(\mathcal{C}=1)$ and that with $\mu < 0$ is trivial $(\mathcal{C}=0)$ and, as such, identified with the vacuum state.

In our analysis of the bulk properties of the PDW state, we shall always be in the regime where $\mu > 0$ (and constant) throughout the system, and consider the effects of a \textit{change in the overall sign of the $p_x+ i p_y$ order parameter.} In this striped system, regions where the order parameter is non-zero 
(regardless of whether it is positive or negative) have the same value Chern number $\mathcal{C}=1$. In spite of that, we shall demonstrate that
the nodes of the order parameter still support gapless modes.
Instead of a single chiral Majorana branch as in the edge of the system discussed in Ref.~\cite{readgreen2000}, 
a node of the PDW order parameter supports rather two non-chiral Majorana branches. 

Below we show that the Lagrangian of the effective low energy theory at each isolated domain wall is
\begin{equation}
\mathcal{L}_{\textrm{d.w.}} =  
i\,\psi_{R}(\partial_{t} - v\,\partial_y)\psi_{R}
+
i\,\psi_{L}(\partial_{t} + v\,\partial_y)\psi_{L}
\,,
\end{equation}
where $\psi_{L/R}$ represent left/right moving massless Majorana fermions. 
This pair of neutral fermion modes --- whose spectrum is identical to that of the one-dimensional critical quantum Ising model --- owe their existence both to a combination of
mirror and chiral (in Class BDI~\cite{Ryu-2010}) symmetries inherent of the Larkin-Ovchinikov order parameter 
as well as to the $p_x+ip_y$ character of the order parameter. In fact, the chiral $p$-wave nature of the order parameter plays a crucial role
in the stability of the fermion zero modes on the nodes of the order parameter, for an earlier analysis \cite{radzihovsky2011} similar in spirit to ours, but in a rather different context of finite momentum $s$-wave superfluids produced by imbalanced cold Fermi gases, 
has found Caroli-De Gennes-Matricon midgap states supported at an isolated node of the $s$-wave order parameter, 
in contrast to the Majorana zero modes of the $p_x+ i p_y$ PDW state.

We further show that the coupling between the domain wall 
counter-propagating Majorana modes leads to a highly nontrivial fermionic spectrum. In general, the (Majorana) fermionic excitations remain gapless. 
Their energy bands cross at the Fermi level, leading to a two-fold degenerate ``Majorana Fermi surface". The Majorana Fermi surface is of topological origin, and the band crossing is protected by a combination of particle-hole symmetry and inversion symmetry~\cite{Agterberg-2017}. Again, the inversion symmetry here crucially relies on both the $p$-wave character of the local pairing and the Larkin-Ovchinikov order parameter.
For PDW states in general, one expects a gapless fermionic spectrum, as a weak PDW order parameter opens gaps only at selected points in $k$-space. In those cases the excitations form a ``Bogoliubov Fermi surface (pocket)", which are closely tied to the original normal state Fermi surface.  Along the Bogoliubov Fermi surface the quasiparticles alternate from being more electron-like to more hole-like. Here we stress that the Majorana Fermi surface is distinct from the original normal state Fermi surface, and satisfy the Majorana condition $\gamma^\dagger (-\vec k) = \gamma(\vec k)$ everywhere. Moreover, in particular ranges of the PDW order parameter, the fermionic spectrum becomes gapped. Interestingly the topology of these gapped phases are distinct from a uniform $p_x+ip_y$ state with a Chern number $\mathcal{C}=1$. Instead we have found phases with both $\mathcal{C}=2$ and $\mathcal{C}=0$, even though locally the pairing is identical to a $p_x+ip_y$ pairing state.

The bulk regions where $\Delta(\vec r)$ is non-zero (which is everywhere except on isolated one dimensional lines extended in the $y$ direction) have the same Chern number and the same the Hall response, irrespective of the overall sign of the order parameter. Consequently, the system is a quantum Hall insulator with respect to the charge modes (albeit with a spatial dependent charge gap) while supporting low energy excitations in the form of gapless neutral fermions supported along the domain walls. 
Thus, while Majorana fermions may tunnel as soft excitations on the PDW domain walls, electron tunneling is suppressed everywhere in the bulk (including along the domain walls) due to the charge gap. The resulting state is an exotic heat conductor but an electric insulator.

Our detailed investigation of the properties of fermionic excitations of the $p_x+ip_y$ PDW state
finds that this system represents a symmetry protected topological phase 
whose remarkably rich properties 
 are summarized as follows:

\paragraph*{1.}Each \emph{isolated} DW supports \textit{a pair} of massless Majorana fermions, as shown in Fig.~\ref{fig: phases}(a), which are protected 
by the unitary symmetry $\mathcal{U} = \mathcal{M}_{y}\,\mathcal{S}$, where $\mathcal{M}_{y}$ is the mirror symmetry along the direction of the domain wall and $\mathcal{S}$ is a chiral symmetry (in class BDI).
In the presence of a uniform component $\Delta_{\textrm{u}}$ of the $p_x+ip_y$-wave order parameter that preserves $\mathcal{U}$ symmetry,
the massless Majorana fermions cannot be gapped out for $|\Delta_{\textrm{u}}| < |\Delta_\textrm{pdw}|$, 
whereas no massless Majorana fermions exist in DWs for $|\Delta_{\textrm{u}}| > |\Delta_\textrm{pdw}|$, representing the phase adiabatically connected to the uniform $p_x+ip_y$-wave state.~\cite{readgreen2000} 

\paragraph*{2.} For $\Delta_{\rm pdw} <v_F$, where $v_F$ is the Fermi velocity of the composite Fermi liquid, in general there exists a two-fold degenerate Majorana Fermi surface (made out of Majorana fermions), protected by the particle-hole symmetry and the inversion symmetry of the PDW state.
As stated above, this state supports gapless neutral excitations but is an electric insulator. This state is one of the main findings of the present work, and we illustrate this phase in Fig.~\ref{fig: phases}(b).

As $\Delta_{\rm pdw}$ varies, this Majorana Fermi surface shrinks and expands periodically, and when the Majorana Fermi surface shrinks to zero size, the fermionic spectrum gets gapped. We found that this gapped state has a Chern number $\mathcal{C}=2$ even if the local pairing is of $p_x+ip_y$ form. 
This can be understood as the result of a Chern-number-one contribution from the bulk $p_x+ip_y$ pairing order in 
addition to a Chern-number-one contribution from the domain walls.
The corresponding quantum Hall state has Abelian topological order, as the vortices of the pairing order do not host Majorana zero modes. The edge conformal field theory (CFT) consists
 a  charge mode and two Majorana fermions, which in total has a chiral central charge $c=2$. This phase is illustrated in Fig.~\ref{fig: phases}(c).

\paragraph*{3.} For PDW states with $\Delta_{\rm pdw}>v_F$, the fermionic spectrum is  gapped (see Fig.~\ref{fig: phases}(d)). From the fermionic point of view, this gapped phase is topologically trivial with $\mathcal{C}=0$
as it does not support a chiral edge Majorana fermions. In the QH setting, we identify this phase with the striped Halperin Abelian
quantum Hall state where electrons form tightly bound charge-2$e$ bosons that condense in a striped Laughlin state.

\paragraph*{4.} The bulk spectrum changes in the presence of a uniform component $\Delta_{\rm u}$ of the $p_x+ip_y$ pairing order. For $\Delta_{\rm pdw}<v_F$, the Majorana FS becomes gapped by an infinitesimal $\Delta_{\rm pdw}$, while for $\Delta_{\rm pdw}>v_F$ the trivial gapped phase survives until a critical value of $\Delta_{\rm u}$. We have found that the gapped phase with $\Delta_{\rm u}$ has a Chern number $\mathcal{C}=1$, i.e., is in the same phase as the uniform Moore-Read $p_x+ip_y$ state. This phase is represented in Fig.~\ref{fig: phases} (e).
 So, interestingly, the neutral FS in Fig.~\ref{fig: phases}(b) represents a quantum critical ``phase" that separates distinct neutral fermion edge states.

Based on our detailed analysis in the remainder of the paper, all these phases mentioned above has been placed in a schematic mean-field phase diagram, shown in Fig.~\ref{fig: phase-diagram}.

\section{Fermionic spectrum of the $p_x+ip_y$ Pair Density Wave}
\label{sec:Properties-Fermionic}

The quantum Hall state with a half-filled Landau level can be viewed as the paired state of the composite fermions coupled to both a dynamical gauge field and the external electromagnetic field. In this section, we analyze the spectrum of the fermionic sector described by the mean-field pairing of composite fermions. We postpone a full description of the quantum Hall state with gauge fields and charge modes to Section~\ref{sec:PDW-FQH}.

 The analysis in this section  also serves as a self-contained mean-field theory for the $p_x+ip_y$ PDW superconductor, which could potentially be relevant for, e.g., Sr$_2$RuO$_4$~\cite{Taillefer-2017}, or superfluid $^3$He~\cite{Davis-Packard-1999}. To our knowledge this theory has not been presented before in the literature.

\subsection{BdG description of the $p_x+ip_y$ PDW state} 
\label{sec:BdG}

Before turning to a PDW state, we consider a generic two-dimensional state with $p_x+ip_y$ local pairing symmetry. We begin with the Bogoliubov-de Gennes (BdG) Hamiltonian in the continuum
\begin{equation}
\label{eq: BdG H}
H(\vec r) = 
\begin{pmatrix}
\epsilon(\vec{k})  & \frac{1}{2} \{  k_{-}, \Delta(\vec r) \}
\\
\frac{1}{2} \{  k_{+}, \Delta^{*}(\vec r) \} & -\epsilon(\vec{k})
\end{pmatrix}	
\,,
\end{equation}
where $\vec{k} = (k_x,k_y) = (-i \partial_x , -i \partial_y)$, $k_{\pm} = k_x \pm i k_y = -i\partial_{\pm}$ (we set $\hbar = 1$). 
{For now let us take the simplest Galilean invariant continuum dispersion  
\be
\epsilon(\vec{k}) = \frac{\vec{k}^{2}}{2m} -\mu.
\ee
We will later discuss the lattice effects of the BdG Hamiltonian.
Here, the anti-commutator
$\{k_-,\Delta(\vec r)\} \equiv k_-\Delta (\vec r)+ \Delta(\vec r) k_-$ is taken to symmetrize the $\vec r$ dependence and $\vec{p}$ dependence, a standard procedure to treat a non-uniform order parameter $\Delta(\vec r)$. 

Throughout this work, we consider the case with a normal-state FS, i.e., $\mu > 0$,
which, in the case of a uniform order parameter $\Delta$, corresponds to the 
``weak-pairing regime", describing a topological paired state with chiral Majorana fermion edge states \cite{readgreen2000}. Notice that the name ``strong-pairing regime" has been used by Read and Green \cite{readgreen2000} for cases with $\mu<0$. Even though we will consider cases with a large pairing order $|\Delta|$, it should not to be confused with the ``strong-pairing regime" in the sense of Read and Green.

The BdG Hamiltonian of Eq.\eqref{eq: BdG H} possesses a particle-hole symmetry
\be
\sigma_1 H \sigma_1 = - H^{*},
\ee	
which relates positive and negative energy states: if $\Psi_E (\vec{r}) = \langle \vec{r} | \Psi_{E} \rangle$ 
is an eigenmode
of $H$ with energy $E$, then $\sigma_1 \Psi^{*}_{E}$ is an eigenmode with energy $-E$. 
Of these states, a particularly interesting eigenstate is the  zero mode (ZM), with $E=0$. It satisfies $\sigma_{1} \Psi^{*}_{0} = \pm \Psi_{0}$ such that they can be expressed as
{$\Psi_{0}(x) = e^{-i\pi/4}\,\psi(x)(1, \pm i )^{T}$}, with $\psi(x) \in \mathbb{R}$.

For a PDW the order parameter varies along the $x$ axis, $\Delta(x)$, and we will work in the gauge where it is a real function of $x$. 
With the ansatz that the zero modes are translation invariant along the $y$ direction ($k_y=0$), the equation for the potential zero modes reads
\begin{equation}
\label{eq: ZM eq v1}
\left( - \frac{\partial^{2}_{x}}{2m} - \mu  \right) \psi(x) 
\pm
\frac{1}{2}\,
\{
\partial_{x},\Delta(x)
\}
\,
\psi(x) 
= 
0.
\end{equation}

It should be emphasized that these states are zero modes of the BdG Hamiltonian, and, as a result, they obey the Majorana condition. 
However, we will see in Sec.~\ref{sec:Domain-wall-bound-states} that here these are not isolated states in the spectrum, 
but are actually part of a branch of propagating massless Majorana fermions, propagating along the  domain wall. 
Thus they should not be confused with their formal cousins, the isolated zero modes at endpoints of one-dimensional $p$-wave 
superconductors \cite{Kitaev-2001}, or at the core of vortices of 2D chiral superconductors \cite{Ivanov-2001}. 
The latter type of zero modes are associated with the non-abelian statistics of these defects, 
whereas the massless Majorana fermions we find here are bound states of domain walls, and are not associated with non-abelian statics. 
For these reasons, and to avoid confusion, we will not refer to the zero modes of the BdG Hamiltonian for  domain walls 
as ``Majorana zero modes.''

\subsection{Domain wall bound states}
\label{sec:Domain-wall-bound-states}

A PDW state is characterized by pairing order parameters $\Delta_{\pm \vec Q}$ (and their higher harmonics such as $\Delta_{\pm 3\vec Q}$, $\Delta_{\pm 5\vec Q}$, ...) with nonzero momentum $\pm \vec Q$, which couple to fermions via
\begin{align}
\!\!H_{\rm pdw} =& \sum_{\vec k, \pm, n=\mathrm{odd}}\Delta_{\pm n\vec Q}f_n({\vec k})\nonumber\\
&\times c^\dagger(\vec k \pm n \vec Q/2) c^\dagger(-\vec k \pm n\vec Q/2)+ \textrm{h.c.},
\end{align}
where $c^{\dagger}(\vec{k})$ is a spinless fermion creation operator at momentum $\vec{k}$, and $f(\vec k)$ is the PDW form factor that is an odd function enforced by fermionic statistics.  At the level of mean field theory, the PDW order parameters $\Delta_{\pm \vec Q}$ satisfy 
\be
|\Delta_{\vec Q}| = |\Delta_{-\vec Q}|,
\label{eq: inversion}
\ee
and this relation holds similarly for all higher harmonics. 
Then the real space form of the order parameter is 
\begin{equation}
\Delta(\vec{r}) = {\sum_{n > 0} |\Delta_{n{\bm Q}}|} {e^{i\theta_{{\bm Q}n}} }
\cos{(n {\bm Q}\cdot {\bm r} + {\phi_{{\bm Q}n}})},
\end{equation}
{At the mean field level, and in the absence of topological singularities, the phases  $\theta_{{\bm Q}n}$ and $\phi_{{\bm Q}n}$ can be both set to zero after a gauge transformation and a spatial translation.}
As we shall see later, this defining property of PDW leads to important symmetries that protect a gapless fermionic spectrum. However, fluctuations about the mean field state do not obey these constraints. As a result, the full PDW order parameter has, in its simplest form, {\em two complex order parameters} , $\Delta_{\pm {\bm Q}}$ \cite{Berg-2008,Berg-2009,Agterberg-2008}. This  complexity of the order parameter manifold has important consequences for the pathways to the quantum and/or thermal melting this state.

In real space, a PDW state can be viewed as a periodic arrangement of domains of pairing order with alternating signs of the order parameter. Across each domain wall the pairing gap $\Delta$ changes sign and vanishes  at the domain wall location. Thus, we expect the low-energy fermionic states to be concentrated to the close vicinity of the domain walls. For simplicity we will only consider the domain wall states with lowest energy. The interplay between higher-energy domain wall states can be similarly analyzed and does not lead to any qualitative differences, as we shall see later. Moreover, it turns out that, for an isolated domain wall, the lowest energy states have interesting topological properties. 

It is convenient to consider a simple picture of a PDW whose $p_x+ip_y$ order parameter has constant magnitude but alternating signs. In this simple case, the midgap states with non-zero energies are pushed away from $E=0$ and we can study the properties of Majorana zero modes more clearly.
We begin our analysis with a single isolated domain wall (DW), or anti-domain wall (ADW), and use the result as a starting point to couple the bound states for a DW-ADW array. It should be noted that the zero modes that we will find below arise as bound states of the BdG one-particle Hamiltonian, much in the same way as  Majorana zero modes at the end-point of a $p$-wave superconductor \cite{Kitaev-2001} (or in the  cores of a half-vortex of a $p_x+ip_y$ superconductor \cite{Ivanov-2001}.) As we noted above, their physics is very different.

We begin with a DW configuration at $x=0$, given by
\begin{equation}
\label{eq: domain wall}
\Delta_{\textrm{isol}}({\vec r}) = - \Delta_{\textrm{pdw}} \,\textrm{sgn}(x)
\,,
\end{equation}
where  $\Delta_{\textrm{pdw}}>0$. {For convenience, we define a quantity with units of momentum
\be
q= m \Delta_{\rm pdw}.
\label{eq:q}
\ee
The solutions to Eq. \eqref{eq: ZM eq v1} and Eq.\eqref{eq: domain wall} yield a pair of normalizable zero-energy solution with $k_y=0$ localized at $x=0${, with even and odd parity}, given by (for more details see Appendix \ref{app:nodes})
\begin{equation}
\label{eq: ZM type 1}
\begin{split}
&\,
\langle\vec{r}|\Psi_{e}\rangle
= 
\frac{N_{e}}{\sqrt{L}}\, e^{-q\,|x|}\,\cos{(\kappa\,x)} \,u_1
\,,
\\
&\,
\langle\vec{r}|\Psi_{o}\rangle
= 
\frac{N_{o}}{\sqrt{L}}\, e^{-q\,|x|}\,\sin{(\kappa\,x)} \,u_1
\,,
\end{split}
\end{equation}
where 
\be
u_1=(1, i)^T/\sqrt{2},~~~
{\kappa} = \sqrt{k_F^2-q^2},
\ee
{$k_F\equiv \sqrt{2m\mu}$ is the Fermi momentum, and the normalization constants $N_{e}, N_{o}$ are given by
\begin{align}
\label{eq:neno}
N_e = \sqrt{\frac{2q(\kappa^2 + q^2)}{\kappa^2 + 2\,q^2}},~
N_o = \sqrt{  \frac{2q(\kappa^2 + q^2)}{\kappa^2}},
\end{align}
where $L$ is the system length along $y$ direction. For $q\ll k_F$ we have $N_o=N_e$, but in general they are different.}

{Notice that the above expression \eqref{eq: ZM type 1} applies to both $q<k_F$ and $q>k_F$: in particular for $q>k_F$ the coefficient $\kappa$ is imaginary and $\cos(\kappa x)$ and $\sin(\kappa x)$ functions in \eqref{eq: ZM type 1} become $\cosh(|\kappa| x)$ and $-i\sinh(|\kappa| x)$ and are non-oscillatory. One can easily verify that the wave functions are still normalizable, thanks to the $e^{-q|x|}$ factor, with the same normalization factor $N_{e,o}$. (Note that $N_o$ becomes imaginary, and $\langle\vec{r}|\Psi_{o}\rangle$ remains real.) However, as we will see, the different forms of the wave packets for $q<k_F$ and $q>k_F$ generally lead to very different coupling between the domain wall modes.}

The dispersion relation of the propagating modes along $y$ axis can be obtained using degenerate perturbation theory by computing the $2 \times 2$
perturbation matrix 
$
\left[\hat{V}(k_y)\right]_{p,p'}
=
\langle\,\Psi_{p}\,|\,\delta H({k_{y}})\,|\,\Psi_{p'}\,\rangle 
$,
for $p,p' = e,o$ and
\be
\delta H({k_{y}})
=
H({k_y}) - H({k_y=0})
=
k_y\,\Delta_{\textrm{isol}}(x) \sigma_y + \frac{k^{2}_{y}}{2m} \sigma_z.
\ee
Direct calculation gives that the eigenstates are a pair of counterpropagating modes:
{\be
\label{eq:LRmodes}
\langle\vec{r}\ket{\Psi_{R,L}(k_y)} = e^{ik_y y} 
\left( \langle\vec{r}\ket{\Psi_{e}} \mp \langle\vec{r}\ket{\Psi_{o}}  \right)/\sqrt{2}
\ee}
with linear dispersion
\begin{align}
E_{R,L}= \pm v_y k_y,~~~v_y = \frac{q^2 }{m\sqrt{2q^2 + \kappa^{2}}}.
\end{align} 
Notice that the quadratic
dependence on the momentum disappears due to $\bra{u_1} \sigma_{z} \ket{u_1} = 0$.

For a ADW configuration with 
\begin{equation}
\label{eq: domain wall}
\Delta'_{\textrm{isol}}({\vec r}) =  \Delta_{\textrm{pdw}} \,\textrm{sgn}(x)
\,,
\end{equation}
the counter-propagating edge states can be straightforwardly obtained by the same procedure. Since a DW and an ADW transforms into each other under a  gauge transformation $\Delta \to -\Delta$, much of the result above for a DW should hold for an ADW. The only difference is that the spinor part $u$ of the wave functions in Eq.\ \eqref{eq: ZM type 1} is replaced with
\be
u_2 = (1, -i)^T/\sqrt{2}.
\ee

{\subsubsection{Symmetry-protected stability of the domain wall counter-propagating modes}}  
\label{sec:symmetry-protection}

The existence of two gapless modes at the domain wall may seem surprising at first sight. After all, a domain wall separates regions with $p_x+ip_y$ pairing and $-(p_x+ip_y)$ pairing, and the two regions have the \emph{same} Chern number. Thus without additional symmetry, the domain wall states are generally gapped.

To establish the stability of the domain wall modes, it is convenient to ``fold" the system along a single domain wall and treat the domain wall as the edge of the folded system. The symmetry that is pertinent to the stability of the edge modes involves a spinless time-reversal $\mathcal{T}=K$ operation ($K$ is the complex conjugation operator). For a  $p_x+ip_y$ state, both the (spinless) time-reversal symmetry $\mathcal{T}$ and the mirror symmetries $\mathcal{M}_{x,y}$ are broken, but one can define a  composite symmetry $\mathcal{M}_{x,y}\mathcal{T}$ that remains intact. Together with the particle-hole symmetry $\mathcal{C}=\tau_x K$ that comes with the BdG Hamiltonian, our (folded) system has a $\mathcal{M}_y\mathcal{S}$ symmetry, where $\mathcal{S}=\mathcal{C}\mathcal{T}=\tau_x$ is known as a chiral operation \cite{Ryu-2010}. The system satisfies
\be
(\mathcal{M}_y\mathcal{S})\mathcal{H}(\mathcal{M}_y\mathcal{S})^{-1}=-\mathcal{H}.
\ee
For the mirror invariant value $k_y=0$, the composite symmetry reduces to a chiral symmetry $\mathcal{S}$, and the 1d subsystem belongs to the BDI class \cite{Ryu-2010}. According to the classification table, BDI class in one dimension  has a $\mathbb{Z}$ classification characterized by an integer winding number $\nu$. We find that the folded system has $\nu=2$, and this corresponds to the two zero modes at the edge at $k_y=0$. One can show that a term $\sim \Delta' \sigma_y$ added to the Hamiltonian of Eq.~\eqref{eq: BdG H} would gap out these two modes, but such a term is prohibited by the $\mathcal{M}_y\mathcal{S}$ symmetry.

We note that the chiral symmetry stems from the defining symmetry of the PDW state. In general, nonuniform superconducting states consists of finite-momentum pairing order parameters $\Delta_{\vec Q}$ and $\Delta_{-\vec Q}$, which are related by inversion. 
The Fulde-Ferrel state, for which $\Delta_{\vec Q}\neq 0$ and $\Delta_{-\vec Q}=0$, 
has a single order parameter and does not oscillate in space.
This SC order parameter in real space has a ``spiral'' pattern in phase rather than the oscillatory pattern. In these cases the $\mathcal{M}_y \mathcal{T}$ symmetry is absent, and so is the $\mathcal{M}_y\mathcal{S}$ symmetry, and there are no such gapless domain wall modes. It is crucial that for a PDW state, similar to a Larkin-Ovchinikov (LO) state, $|\Delta_{\vec Q}|=|\Delta_{-\vec Q}|$, such that the $\mathcal{M}_y\mathcal{S}$ symmetry is intact.

{\subsection{FS from domain wall coupling}} 
\label{sec:fs-rec}

So far we have considered the case of completely isolated DWs. At finite values of the PDW wavelength, though, hybridization between DWs inevitably occurs, and is responsible
for making the DW excitations regain their 2D character. {In this Subsection, we will consider a PDW state with DW (and ADW) bound states and derive the dispersion of the (hybridized) bulk states.}

Due to the exponential decay of the domain wall state wave function in Eq.\eqref{eq: ZM type 1}, we expect the effective hopping matrix elements 
between DWs separated by distance $d$ scales as $e^{-qd}$
and, for nearest neighbor DW and anti-DW separated by $\lambda/2 \equiv \pi/Q$, the coupling is of the order $e^{-\pi q/Q}$. 
Then if $Q<q$,
we can employ a tight-binding approximation where the nearest neighbor hopping gives the dominant contribution.

For the rest of the work, we will mainly focus on the regime 
\be
Q<q<k_F,
\ee
where the first inequality enables us to use a tight-binding approximation, and the second inequality ensures that the local pairing gap 
$\Delta(\vec r)$ is smaller than Fermi energy $\mu$, a reasonable assumption {in the spirit of the weak coupling 
 theory}. As we discussed in Sec.\ \ref{sec:Domain-wall-bound-states}, in this regime the wave functions in \eqref{eq: ZM type 1} are oscillatory functions enveloped by symmetric exponential decay.  We have set the PDW wavevector $Q<k_F$ --- this is needed in order for the normal state FS to be reconstructed in a meaningful way. As we proceed, we will discuss other regimes of the length scales with $q<Q$ and $q>k_F$ as well.
 
Consider the PDW state obtained as a periodic sequence of DWs and ADWs,
\begin{equation}
\label{eq: PDW order parameter - case 1}
\begin{split}
&\,
\Delta(x) 
=
\Delta_{\rm pdw} \Big[ 1 + \sum_{\epsilon = 1,2} \sum_{n \in \mathbb{Z}}  
(-1)^{\epsilon}\,\textrm{sgn}(x - x^{(\epsilon)}_{n}) \Big]
\,,
\end{split}
\end{equation}
where DWs are located at $x^{(1)}_n =  n\,\lambda$ and 
ADWs are located at $x^{(2)}_{n} = (n+1/2)\,\lambda$.
The order parameter of \eqref{eq: PDW order parameter - case 1}, and consequently 
the BdG Hamiltonian of the state,  are then periodic under shifts of $x$ by integer multiples
of $\lambda$. 
{The conclusion of this paper does not change if one uses a smooth order parameter instead of the square-wave one. The main reason is that the essential ingredient for our result is the counter-propagating Majorana modes at an isolated domain wall and the two-fold degenerate Majorana FS via domain wall coupling. Both are protected by symmetry, shown in Sec.~\ref{sec:symmetry-protection}, and are not affected by the detailed form of the PDW order parameter. We adopt this square-wave simply to simplify the analytical calculation.}

Other than this translational symmetry, the PDW configuration \eqref{eq: PDW order parameter - case 1} also entails an inversion symmetry of the BdG Hamiltonian \eqref{eq: BdG H} with inversion centers at $x_n^{(\epsilon)}$. Indeed, under such an inversion, both $k_{\pm}$ and $\Delta(\vec r)$ change sign, rendering their anticommutator and hence $H(\vec r)$ invariant. For the domain wall modes, from Eq.~\eqref{eq:LRmodes} we see that left movers and right movers transform into each other under inversion. It is straightforward to see that this inversion symmetry simply derives from Eq.~\eqref{eq: inversion}, the defining property of a PDW state.

The system also has a ``half-translation" symmetry. Namely, under a translation by $\lambda/2$ the order parameter \eqref{eq: PDW order parameter - case 1} flips sign, but this is identical to the original state after a gauge transformation $\Delta\to -\Delta$.  For the domain wall modes, left and right movers retain their chirality under the half translation.
We will use these symmetries to establish relations between the hopping matrices.

Let us consider a variational state
\begin{equation}
\label{eq: Bloch state}
\ket{\Psi_{k_{x},k_{y}}}
=
\sum_{\epsilon = 1,2} \sum_{\mu = L,R}
c_{\epsilon,\mu}
\sum_{n \in \mathbb{Z}}
\frac{e^{i k_{x}\,n\,\lambda}}{\sqrt{N}} 
\ket{\Psi_{\epsilon,\mu,n}(k_y)}
\,,
\end{equation}
where the subscript $\epsilon =1$ denotes DW modes and $\epsilon=2$ ADW modes. 
Recall that $\ket{\Psi_{\epsilon=1}}\propto u_1 = (1,i)^T/\sqrt{2}$ and $\ket{\Psi_{\epsilon=2}}\propto u_2 =(1,-i)^T/\sqrt{2}$.
The coefficients $c_{\epsilon,\mu}$ are variational parameters,
the dependence on the momentum $k_y$ enters via the dispersive modes $\ket{\Psi_{L,R}}$ along
each DWs and ADWs and the dependence on the crystal momentum $k_x\in(-Q/2, Q/2)$ enforces that the state of Eq. \eqref{eq: Bloch state}
satisfies the Bloch theorem. 

\begin{figure}
\includegraphics[width=\columnwidth]{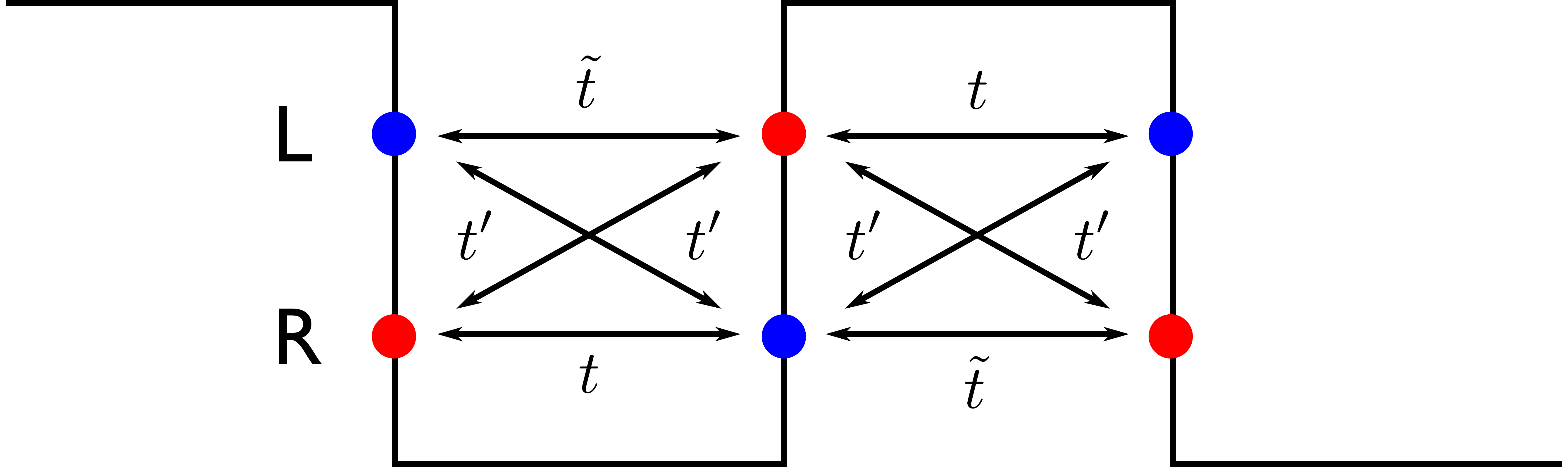}
\caption{The coupling $t$, $\tilde t$, and $t'$ between the neighboring domain wall modes.}
\label{fig:tibi}
\end{figure}

The steps leading to the energy of this variational tight-binding state are lengthy
but straightforward \cite{ziman-book}, and are presented in Appendix~\ref{app:tight-binding}.
Minimization of the  
energy of the state 
\begin{equation}
E_{k_{x},k_{y}}[\{ c_{\epsilon, \mu} \}] = \dfrac{\bra{\Psi_{k_{x},k_{y}}} H \ket{\Psi_{k_{x},k_{y}}}}
{\bra{\Psi_{k_{x},k_{y}}} \Psi_{k_{x},k_{y}} \rangle},
\end{equation}
with respect to the variational parameters $\{ c_{\epsilon, \mu} \}$,
yields the secular equation
\begin{equation}
\label{eq: effective H 4x4}
\textrm{det}
\left[
\mathcal{H}_{k_{x},k_{y}}
-
E_{k_{x},k_{y}}
\,
\mathrm{I}_{4 \times 4}
\right]
=
0,
\end{equation}
where the effective Hamiltonian, {valid in the vicinity of $k_y=0$,} is given by
\begin{align}
&\mathcal{H}(k_x,k_y)
=\nonumber\\
&\begin{pmatrix}
v_{y}k_y
& 
0 
& 
t' + t'e^{-ik_x} 
&
t- \tilde{t}e^{-ik_x} 
\\
0 
&
-v_{y}k_y
&
-\tilde{t} + te^{-ik_x} 
&
t' + t'e^{-ik_x} 
\\
t' + t'e^{ik_x} 
&
-\tilde{t} + te^{ik_x} 
&
v_{y}k_y
&
0
\\
t - \tilde{t}e^{ik_x} 
&
t' + t'e^{ik_x} 
&
0
&
-v_{y}k_y
\end{pmatrix}
,
\label{eq: tight-binding H}
\end{align}
where for convenience we have redefined $ k_x\lambda \to k_x$ so that $k_x\in (-\pi, \pi)$.

This effective Hamiltonian is expressed in the basis of states 
$\{ \ket{\Psi_{1,R}},\ket{\Psi_{1,L}},\ket{\Psi_{2,R}},\ket{\Psi_{2,L}} \}$ (momentum dependence omitted),
where the indices $1$($2$) denote DW (ADW) degrees of freedom. The diagonal blocks proportional to $v_{y}k_{y}\sigma_3$ then represent
the kinetic energies of the right- and left-moving modes on DWs and ADWs, respectively, while the off-diagonal blocks
represent the coupling between an adjacent DW-ADW pair. 
The constants $t, \tilde {t}, t'$ can be understood intuitively as ``hopping amplitudes" between the neighboring domain wall modes, which we illustrate in Fig.~\ref{fig:tibi}. Specifically, $t'$ describes the coupling between neighboring modes with the \emph{same} chirality. Importantly, {all} these couplings are the same following the inversion symmetry and the half-translation symmetry. $t$ describes the coupling between the right-mover at a DW with the left-mover at an ADW \emph{to its right}. By the half-translational symmetry or inversion symmetry,  $t$ also describes the coupling between the right-mover at a ADW with the left-mover at an DW \emph{to its right}. On the other hand, $\tilde t$ describes the coupling of a left-mover with a right-mover \emph{to its left}. Notice that there are no symmetry requirement relating $t$ and $\tilde t$.

In Appendix \ref{app:tight-binding}, we evaluated $t,\tilde t$, and $t'$, and the results are,
\begin{align}
\label{eq:ts}
t = -&\frac{\kappa }{4m}\exp{(-q\lambda/2)}\nonumber\\&\times\left[2N_eN_o\cos\left (\frac{\kappa\lambda}{2}\right)+(N_e^2-N_o^2)\sin\left(\frac{\kappa\lambda}{2}\right)\right] \nonumber\\
\tilde{t} = -&\frac{\kappa }{4m}\exp{(-q\lambda/2)}\nonumber\\&\times \left[2N_eN_o\cos\left (\frac{\kappa\lambda}{2}\right)-(N_e^2-N_o^2)\sin\left(\frac{\kappa\lambda}{2}\right)\right] \nonumber\\
\!\!\!t'= -&\frac{\kappa }{4m}\exp{(-q\lambda/2)}(N_e^2+N_o^2)\sin\left(\frac{\kappa\lambda}{2}\right)
\end{align}

{{We note that so far our analysis and Eqs.~(\ref{eq: tight-binding H}, \ref{eq:ts}) apply to \emph{both} $q<k_F$ \emph{and} $q>k_F$. In particular, it is easy to verify that for $q>k_F$, $t$, $\tilde t$, and $t'$ are still real.
As we promised, we will focus on  $q<k_F$ for now.}} In this regime, we find that out of the four bands~\footnote{For the rest of the work, we use ``band'' and ``band topology'' to refer to those for the BdG Hamiltonian} given by Eq.~\eqref{eq: tight-binding H}, two of them cross each other at zero energy, illustrated in Fig.~\ref{fig:bc} the eigenstate of \eqref{eq: tight-binding H} for $t'=0.5$, $t=0.4$, $\tilde t=0.6$ at $k_y=0$.
\begin{figure}
\includegraphics[width=.8\columnwidth]{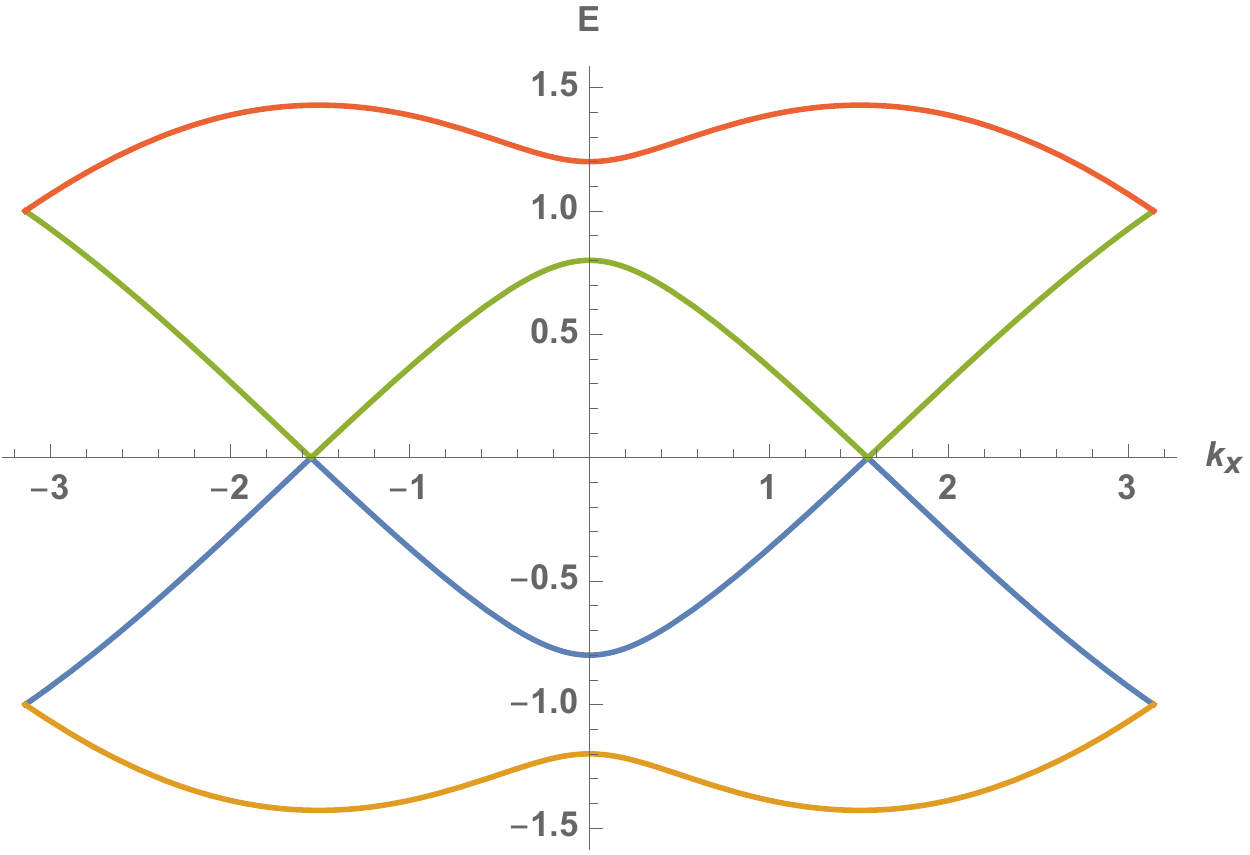}
\caption{The energy bands of the tight-binding Hamiltonian \eqref{eq: tight-binding H} for $t'=0.5$, $t=0.4$, $\tilde t=0.6$ at $k_y=0$.}
\label{fig:bc}
\end{figure}
The zero-energy band crossing results in a (\emph{two-fold degenerate}) FS, whose contour is given by the vanishing of the determinant
\begin{align}
\label{eq:fs}
&\det[\mathcal{H}(k_x,k_y)]\nonumber\\
&=\left[ (v_y k_y)^2 - 4(t'^2+t\tilde t) \cos^2(k_x/2)  + (t+\tilde t\,)^2\right ]^2 = 0.
\end{align}
It is easy to verify that this equation does have a solution for $q<k_F$. Importantly, this degenerate FS belongs to energy bands of \emph{Majorana modes}, and by construction quasiparticles near it satisfy the Majorana condition $\gamma^\dagger(\vec k)=\gamma (-\vec k)$. For this reason, we term it a ``Majorana FS".

To verify Eq.~\eqref{eq:fs}, we numerically solved the lattice version of the BdG Hamiltonian \eqref{eq: BdG H}. For the normal state we used the dispersion 
\be
\label{eq:lat1}
\epsilon({\vec k}) = -t_0(\cos k_x + \cos k_y)-\mu,
\ee and for the off-diagonal element of the BdG Hamiltonian we used
\be
\label{eq:lat2}
\begin{split}
&\,
{k_{\pm} \rightarrow \sin  k_x \pm i\sin{ k_y}}
\end{split}
\ee 
We set the parameters as $t_0=1$, $\mu = -1.25$, $\Delta_{\rm pdw} = 0.82$, and $Q=\pi/6$. The match between the computed spectral function $\rho({\vec k}, E)$ and the FS analytically given by \eqref{eq:fs} is good, as shown in Fig.~\ref{fig: analytical FS}. The match becomes even better if we take $\mu \to -2$. In this case the relevant dispersion becomes parabolic and approaches the continuum limit.

As $q\equiv m\Delta_{\rm pdw}$ varies, the relative amplitudes of $t,\tilde t$, and $t'$ varies periodically, and the the two-fold degenerate FS expands and shrinks. Note that at 
\be
\sqrt{k_F^2-q^2}\equiv \kappa=nQ, 
~n\in \mathbb{Z},
\ee
 from Eq.~\eqref{eq:ts} we have $t'=0$ and $t=\tilde t$. Plugging these into Eq.~\eqref{eq:fs} we see that the two-fold degenerate FS shrinks to \emph{two} Dirac points both at $\vec k = 0$. However, we will see in the next Subsection that the existence of two overlapping Dirac points, i.e., the four-fold degeneracy at $\vec k=0$, is 
a non-universal property of the continuum theory,
and in generic cases at $\kappa=nQ$ the fermionic spectrum is actually gapped. To that end, we will first need to understand whether and why the band crossing at the FS for generic values of $t,\tilde t$, and $t'$ is robust.

\begin{figure}
\includegraphics[width=0.49\columnwidth]{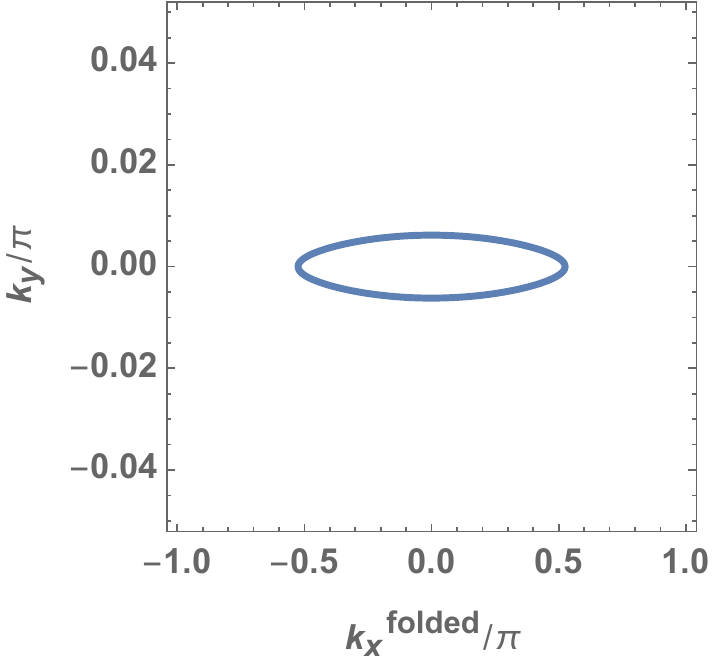}
\includegraphics[width=0.495\columnwidth]{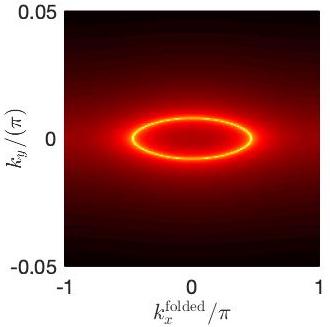}

\caption{FS from coupling the domain wall modes.
Left panel: FS obtained from Eq.~\eqref{eq:fs}, where the hopping parameters in Eq.~\eqref{eq:ts}
were computed for lattice parameters $t_0=1$, $\mu=-1.25$, $\Delta_{\textrm{pdw}}=0.82$ and $Q = \pi/6$.
 Right panel: Simulated fermionic spectral function $\rho(\vec k, E=0)$ from a lattice nearest-neighbor hopping model subject to a PDW order parameter.  }
 \label{fig: analytical FS}
 \end{figure}

 \begin{figure}
 \includegraphics[width=\columnwidth]{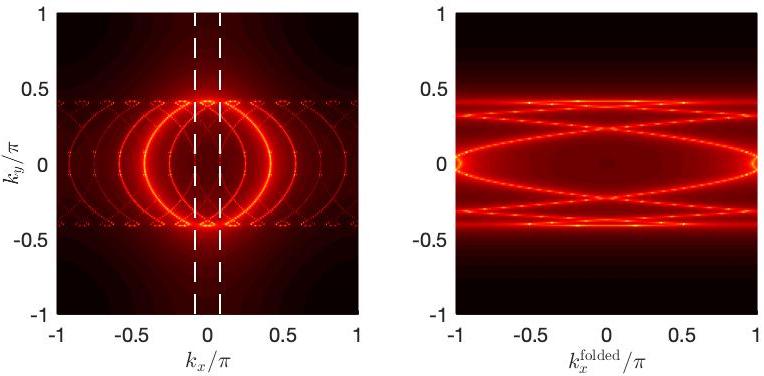}
 \caption{{Simulated fermionic spectral function $\rho(\vec k, E=0)$ from a lattice nearest-neighbor hopping model subject to a PDW order parameter. For the normal state we used the dispersion $\epsilon({\vec k}) = -t_0(\cos k_x + \cos k_y)-\mu$, and we took the local PDW coupling with wave vector $Q$ as $\pm\Delta_{\rm pdw} (\sin k_x + i\sin{k_y})c^\dagger_{\vec k} c^\dagger_{-\vec k}+h.c$. We set the parameters as $t_0=1$, $\mu = -1.25$, $\Delta_{\rm pdw} = 0.052$, and $Q=\pi/6$. In the left panel we plot the spectral function in the original Brillouin zone. In right panel, we plot the spectral function in the folded Brillouin zone (the region between the dashed lines in the left panel) with a new lattice constant $a_Q= a_0\times2\pi/Q$. Compared with Fig.~\ref{fig: analytical FS}, the FS here can be viewed as a small perturbation of the original (circular) FS formed by composite fermions.}}
  \label{fig: numerical FS}
 \end{figure}

Before we move on, let us briefly discuss the fermionic spectrum for $q\ll Q$. So far we have worked in the regime where we only need to consider the nearest-neighbor coupling between the domain wall Majorana modes. For $q\ll Q$, the domain wall states are no longer well-defined, as their localization length becomes longer than the PDW wavelength. In this case the domain wall Majorana modes are not a good starting point for analytical calculations. It turns out this regime admits a simple description in $\bm k$ space. We note that due to Brillouin zone folding, the typical energy scale for the relevant bands in the folded BZ is given by $E_F^Q\sim v_F Q$. In this regime we have $k_F\Delta_{\rm pdw}\ll E_F^Q$, $k_F\Delta_{\rm pdw}$ being the size of the $p$-wave gap on the FS, which indicates that PDW can be treated perturbatively in $\bm{k}$ space. Indeed, numerically we found that the FS resembles that of the composite fermions, except at the regions with $k_x=\pm Q/2,~\pm 3Q/2,~\cdots$, which gets gapped and perturbatively reconstructed by the PDW order,
as shown in Fig.~\ref{fig: numerical FS}.

 Importantly, in this case the FS are made out of Bogoliubov quasiparticles $d(\vec k) = u_{\vec k} c(\vec k) + v_{\vec k}c^\dagger (-\vec k +\vec Q)$, which are in general not Majorana quasiparticles, i.e., $u_{\vec k}\neq v_{\vec k}$. For this reason we call it the ``Bogoliubov FS" to distinguish it from the Majorana FS we obtained previously.
 As one increases $\Delta_{\rm pdw}$, the Bogoliubov FS gets progressively gapped and crosses over to the one obtained previously in Fig.~\ref{fig: analytical FS}.\\

{\subsubsection{Symmetry-protected stability of the Majorana FS}}  
\label{sec:symmetry-protection-fs}

As we emphasized, two bands cross at the FS given by \eqref{eq:fs}. It is then a natural question whether this band crossing is robust against perturbations, or it is accidental due to the particular BdG Hamiltonian \eqref{eq: BdG H} we are using. Here we show that the gapless nature of the FS is protected by symmetry. In particular, the defining inversion symmetry of the PDW state $|\Delta_{\vec Q}| = |\Delta_{-\vec Q}|$ again plays a crucial role.

In the literature, band crossings in $\vec k$ space that form sub-manifolds with co-dimension 2 and 3 have been intensively discussed. In two spatial dimensions  the band crossings are known as Dirac points, while in three spatial dimensions, these are Weyl points (with co-dimension 3), Dirac points (with co-dimension 3), and nodal lines (with co-dimension 2). The band crossing we obtained has co-dimension 1, which corresponds to ``nodal FS's". The stability of the nodal FS is less-well known, but has also been recently analyzed~\cite{Sato-2014,Zhao-Schnyder-Wang-2016,Agterberg-2017,Sigrist-2017,Moroz-2018,Agterberg-2018}. A particularly systematic analysis has been done in Ref.~\cite{Sigrist-2017}.

\begin{figure}
\includegraphics[width=0.8\columnwidth
]{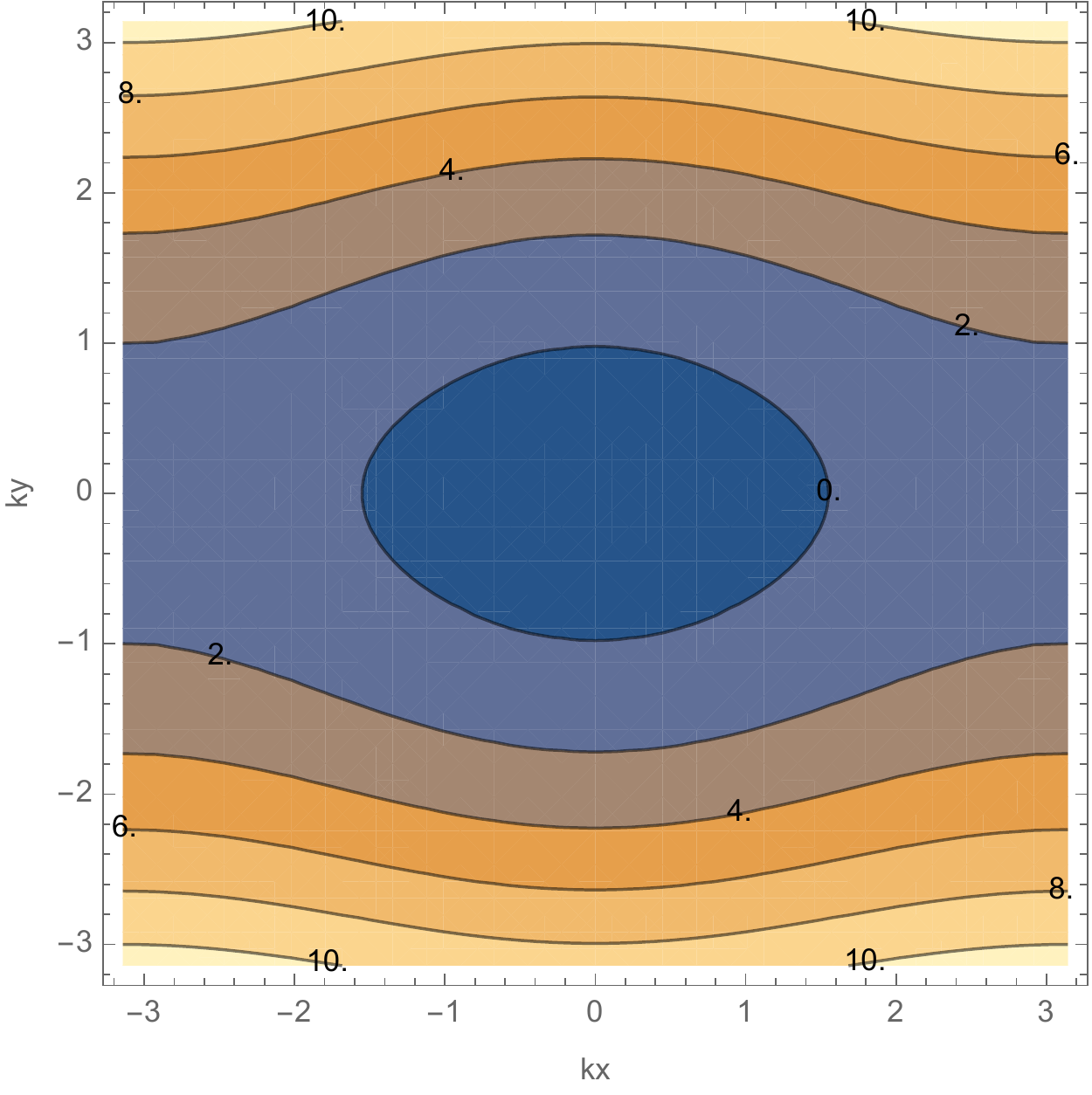}
\caption{Contour plots of the Pfaffian Eq.~\eqref{eq: pfaffian}  $t = 0.4$, $t' = 0.5$, $\tilde{t} = 0.4$ showing the FS for
at $\textrm{Pf}(\tilde{\mathcal{H}}(\vec{k})) = 0$, which separates the region where
the Pfaffian changes sign. For parameter values such that $\kappa = n\,Q$, which correspond to $t'=0$ and $t=\tilde{t}$,
the FS shrinks to a doubly degenerate Dirac point at $(k_x, k_y)=(0,0)$. We note, however, that this Dirac point is accidental, 
in the sense that it is a property of the continuum approximation of the band structure in which the original FS (in the absence of a 
PDW order parameter) is circular. Our numerical calculation indeed shows that this Dirac point is gapped once lattice effects become
non-negligible.  
}
\label{fig: pfaffian}
\end{figure}

For our purposes, we will closely follow the analysis in Ref.~\cite{Agterberg-2017}. We focus on the particle-hole symmetry and the inversion symmetry previously identified for the BdG Hamiltonian \eqref{eq: BdG H} for the PDW state. With regard to the effective Hamiltonian Eq.~\eqref{eq: tight-binding H}, the particle-hole symmetry that relates positive and negative energy states of the Hamiltonian Eq.~\eqref{eq: tight-binding H} is expressed through a {unitary} operator $\mathcal{C}$:
\begin{equation}
\label{eq: PH symmetry}
\begin{split}
&\,
\mathcal{C}\,\mathcal{H}(-\vec{k})^{T}\,\mathcal{C}^{\dagger} = - \mathcal{H}(\vec{k})
\,,
\\
&\,
\mathcal{C} = \sigma_{0}\otimes\tau_{3} = 
\begin{pmatrix}
\sigma_{0} & 0
\\
0 & -\sigma_{0}
\end{pmatrix}
\,.
\end{split}
\end{equation} 
The effective Hamiltonian Eq.~\eqref{eq: tight-binding H} also inherits the inversion symmetry of the PDW state under inversion operation $x \rightarrow -x$
with respect to the domain wall. Because the right- and left moving states Eq.~\eqref{eq:LRmodes} are formed
by linear combinations of even and odd parity states, they are switched under inversion. Inversion
symmetry is represented in the effective theory Eq.~\eqref{eq: tight-binding H} {by  a unitary operator $\mathcal{I}$ satisfying}
\begin{equation}
\label{eq: inversion symmetry}
\begin{split}
&\,
\mathcal{I}(-k_{x})\,\mathcal{H}(-\vec{k})\,\mathcal{I}(-k_x)^{\dagger} =  \mathcal{H}(\vec{k})
\,,
\\
&\,
\mathcal{I}(k_x) = 
\begin{pmatrix}
\sigma_{1} & 0
\\
0 & e^{-i k_x}\,\sigma_{1}
\end{pmatrix}
\,,
\end{split}
\end{equation} 
where the action of $\sigma_1$ is to switch left- and right-moving modes and the momentum dependence
$e^{-i k_x}$ on the ADW degrees of freedom reflects that fact that the center of inversion is taken with respect to a DW. {Both $\mathcal{C}$ and $\mathcal{I}$ relates $\vec k$ with $-\vec k$, and it is useful to consider their composite that relates $\mathcal{H}({\vec k})$ with itself. We define another unitary operator} 
\be
U_{\mathcal{CI}}(k_x)\equiv  \mathcal{C} \mathcal{I}(k_x) = \begin{pmatrix}
\sigma_{1} & 0
\\
0 & -e^{-i k_x}\,\sigma_{1}
\end{pmatrix},
\ee 
{which, importantly, is symmetric. It then follows that for any given $\vec k$,}
\begin{equation}
\label{eq: CI symmetry}
\begin{split}
&\,
U_{\mathcal{C I}}(k_x)\,\mathcal{H}(\vec{k})\,U^{\dagger} _{\mathcal{C I}}(k_x)=  -\mathcal{H}^{T}(\vec{k})
\,.
\end{split}
\end{equation} 
{Any symmetric matrix can be diagonalized as $U_{\mathcal{CI}}=Q\Lambda Q^T$, with $\Lambda$  diagonal and $Q$  unitary.
Then, as shown in Ref.~\cite{Agterberg-2017}, Eq.~\eqref{eq: CI symmetry} can be used to define 
an anti-symmetric
$\tilde{\mathcal{H}}(\vec{k}) = \Omega(\vec{k})\,\mathcal{H}(\vec{k})\,\Omega^{\dagger}(\vec{k})$,
with $\tilde{\mathcal{H}}(\vec{k}) = -\tilde{\mathcal{H}}^{T}(\vec{k})$,
where $\Omega(\vec{k})={\sqrt{\Lambda^\dagger}} Q^\dagger$ is unitary.}
The antisymmetric nature of $\tilde{\mathcal{H}}(\vec{k})$ allows us to express the determinant at any given $\vec k$ in terms 
of the Pfaffian as
\begin{align}
\textrm{det}(\mathcal{H}(\vec{k})) = \textrm{det}(\tilde{\mathcal{H}}(\vec{k})) = \left[\textrm{Pf}(\tilde{\mathcal{H}}(\vec{k}))\right]^2
\,.
\end{align}
At the location of the FS,  $\det(\mathcal{H}(\vec k)) = \textrm{Pf}(\mathcal{H}(\vec k)) = 0$. 
Importantly, since $\mathcal{H}$ is Hermitian, one can check that the Pfaffian $\textrm{Pf}(\tilde{\mathcal{H}}(\vec{k}))$ is always real.  If
two points at the BZ $\vec{k}_1$ and $\vec{k}_2$ satisfy 
$
\textrm{Pf}(\tilde{\mathcal{H}}(\vec{k}_1))
\,
\textrm{Pf}(\tilde{\mathcal{H}}(\vec{k}_2))
<
0
$, 
then there is a FS separating $\vec{k}_1$ and $\vec{k}_2$ at which the Pfaffian changes sign. Symmetry-preserving perturbations can move the location of the FS in $\vec k$ space, but they cannot gap the spectrum unless the FS shrinks to zero size.

Specifically for our tight-binding Hamiltonian \eqref{eq: tight-binding H} one obtains
\be
\textrm{Pf}(\tilde{\mathcal{H}}(\vec{k}))
=
(v_y k_y)^2 + (t+\tilde t\,)^2 - 4(t'^2+t\tilde t) \cos^2(k_x/2).
\label{eq: pfaffian}
\ee
The contour plots of the Pfaffian are plotted in Fig.~\ref{fig: pfaffian}.
The condition $\textrm{Pf}(\mathcal{H}{(\vec k))} = 0$ indeed matches  the location
of the FS given by \eqref{eq:fs}. The FS is stable in the presence
of small perturbations that preserve the two symmetries simultaneously.

{\subsection{ Gapped states from domain wall coupling}} 
\label{subsubsec: Trivial phase}

\subsubsection{gapped phase near $\kappa =n Q$}
\label{subsubsec: c=2}

We continue to focus on the regime $Q<q<k_F$.
Our argument on grounds of inversion symmetry in Sec.~\ref{sec:symmetry-protection-fs} establishes the stability of the doubly degenerate FS. However, it \emph{does not} ensure the stability of the double Dirac points obtained from \eqref{eq:fs} and \eqref{eq:ts} at $\kappa=nQ$, which in turn are obtained from the continuum BdG Hamiltonian Eq.~\eqref{eq: BdG H}. Here we show that for a  BdG Hamiltonian with a generic lattice dispersion and $p$-wave form factor of local pairing, the Dirac spectra at $\kappa = nQ$ in the continuum model are replaced by gapped fermionic spectra. Moreover, remarkably, the gapped system has a trivial band topology, even though the local pairing symmetry is $p_x+ip_y$ with $\mu>0$.

\begin{figure}[h]
\includegraphics[width=\columnwidth]{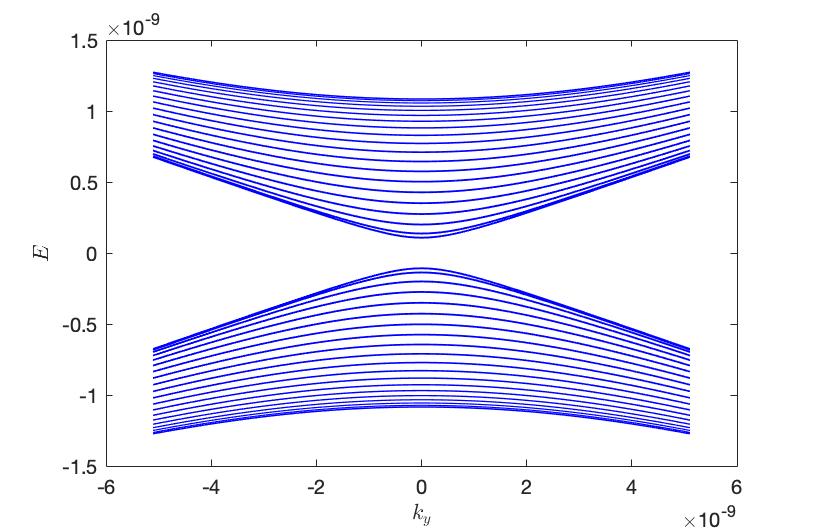}\\
\centering{(a)}\\
\includegraphics[width=\columnwidth]{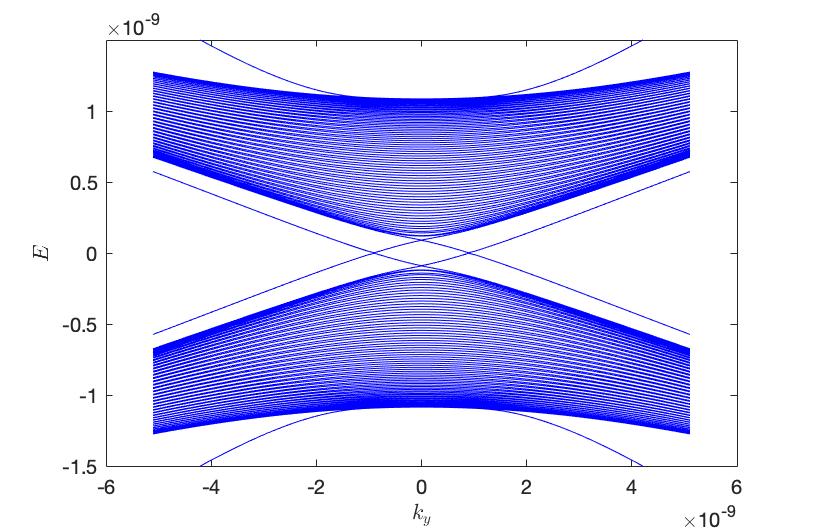}\\
\centering{(b)}
\caption{The gapped Dirac spectrum from a numerical calculation near $\kappa = 8Q$. We have used the lattice model in Eqs.\ \eqref{eq:lat1} and \eqref{eq:lat2}, with $t=1$ and $\mu =1.9$. The chemical potenial is very close to the band bottom, thus the system is quasi-continuous, $k_F \ll 2\pi /a $.
 Panel (a) is obtained by the periodic boundary condition in $x$ direction, while Panel (b) is from the open boundary condition. From counting the number of the localized edge modes in (b), such a phase has a Chern number $\mathcal{C}=2$. {The smallness of the energy eigenvalues comes from the extreme tight-binding limit we take $q\gg Q$, which ensures a quantitative match with analytical results.}}
\label{fig:gapped dirac}
\end{figure}

It is instructive to first understand the origin of the double Dirac points at $\kappa = nQ$ in the continuum model. At these points, from Eq.~\eqref{eq:ts}, the same-chirality hopping amplitude $t'$ vanishes. As a result, a left mover only couple to their adjacent right movers and vise versa. The domain wall modes decompose to two separate chains of coupled wires, each of them alternating between left- and right-movers. We illustrate this situation in Fig.~\ref{trivial phase}, where the solid arrowed lines denote $\tilde t$ and dashed arrowed lines denote $t$. From Eq.~\eqref{eq:ts} we see that at $\kappa = nQ$ we have $t=\tilde t$. With $t=\tilde t$, each of two chains gives rise to a Dirac point at $\vec k=0$, in a mechanism similar to the Dirac cone ``reconstruction" at the surface of a topological insulator
via hybridization of chiral modes localized at oppositely oriented ferromagnetic domain walls.~\cite{Mross2016}. 

However, recall that after a careful analysis we have concluded there is \emph{no} symmetry that relates $t$ and $\tilde t$. The fact that we obtained $t=\tilde t$ in the continuum model at $\kappa = nQ$ is merely an accident. For a generic dispersion with an \emph{almost} circular normal state FS, we expect from \eqref{eq:ts} that when $t'=0$, 
\be
\tilde t- t\propto N_e^2-N_o^2.
\ee
Following an analogy with the well-known Su-Schrieffer-Heeger model for polyacetylene~\cite{Su-1979}, this asymmetrical coupling pattern gaps out the fermionic spectrum.  The spectral gap is proportional to $N_o^2-N_e^2$. This spectral gap is rather small --- in particular for $q\ll k_F$ we have from \eqref{eq:neno} that $N_e\approx N_o$. For $q$ more comparable to $k_F$ this spectral gap increases. Naturally, in the vicinity of the would-be Dirac point values, i.e., near $\kappa=nQ$, the spectral gap persists, and for larger $q$'s, the range of $q$ with a gapped spectrum is larger.

Indeed, we numerically solved the lattice version of the problem with Eqs.\ \eqref{eq:lat1} and \eqref{eq:lat2}. With $t=1$ and $\mu =1.9$, the normal-state FS is nearly isotropic. Yet we see that when the FS shrinks it becomes gapped, instead of Dirac points. We show the gapped Dirac dispersion in this situation in Fig.~\ref{fig:gapped dirac}(a). We have also verified that as $\Delta_{\rm pdw}$ increases, the Dirac gap becomes larger.

\begin{figure}
\includegraphics[width=\columnwidth]{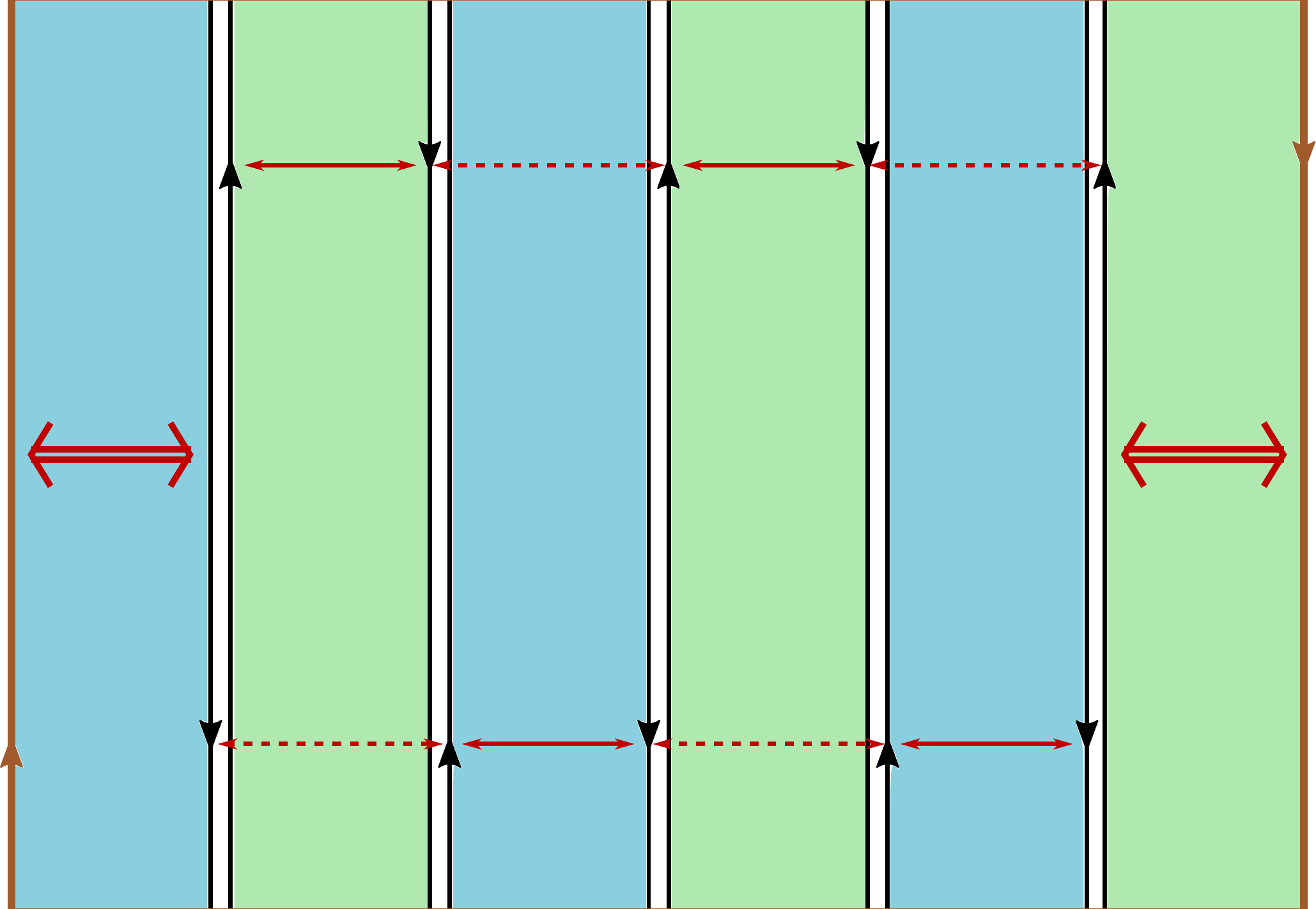}
\caption{Illustration of the gapped fermionic spectrum with Chern number $\mathcal{C}=0$ or $\mathcal{C}=2$, depending on whether $\tilde t$ (solid double lines) or $t$ (dashed double lines) is larger.}
\label{trivial phase}
\end{figure}

The band topology of this gapped phase can be obtained by inspecting the edge modes.
From Fig.~\ref{trivial phase}, it is straightforward to see that the coupling pattern between the domain wall modes (not including the leftmost and rightmost modes, which are edge modes), leaves two unpaired chiral domain wall modes at the two ends. On the other hand, owing to the local $p_x+ip_y$ pairing symmetry, there would be a chiral mode (shown in yellow in Fig.~\ref{trivial phase}) at each physical edge of the system. For $\tilde t> t$, i.e., when the hopping represented by solid arrowed lines is stronger, one can check that the unpaired domain wall mode and the would-be edge mode are of opposite chirality, and they gap each other. The resulting state does not host any gapless edge modes, and is thus topologically trivial with Chern number $\mathcal{C}=0$. On the other hand, if $t>\tilde t$,  the unpaired domain wall mode and the edge mode are of the same chirality; in this case
 at each edge there would be two chiral modes propagating in the same direction, with $\mathcal{C}=2$. A similar situation has been found in a $p$-wave SC in the presence of a vortex lattice~\cite{Vafek-2015}.

{
It is also instructive to understand how the competition between $t$ and $\tilde{t}$ changes the Chern number by $2$,
by considering the following
reasoning. Let $t' =0$, $t = \tau + \delta$ and $\tilde{t} = \tau - \delta$, then the effective Hamiltonian, after an appropriate unitary 
transformation, is in the form $\mathcal{H} = \vec{B}(\vec{k})\cdot\vec{\Gamma}$,
where $\vec{B}(\vec{k}) = (v_{y}k_{y}, -2\tau\sin{(k_x/2)},2\delta\cos{(k_x/2)})$ and
$\vec{\Gamma} = (\Gamma_1, \Gamma_2, \Gamma_3)$ are anti-commuting matrices with $\Gamma^{2}_i=1$.
Then at $\delta=0$ ($t = \tilde{t}$) we see the two Dirac points at $\vec{k}=0$, which become massive
for $\delta\neq 0$. The Chern number measures the winding of the spinor $\vec{B}(\vec{k})$ as $\vec{k}$
is varied. Importantly, the sign of $\delta$ controls the orientation of the spinor along
the third axis (i.e. direction $\Gamma_3$). Reversing the sign of $\delta$ reverses the 
orientation of the spinor and changes the Chern number by $\Delta C = 2 \times 1 = 2$,
where the factor of $2$ accounts for the number Dirac cones.}
 
Notice that for both $\mathcal{C}=0$ and $\mathcal{C}=2$ pairing  states there are  \emph{no} Majorana zero modes bound at vortex cores. In particular for $\mathcal{C}=2$ state there are two would-be zero modes near a vortex  that generally gap each other. In our PDW setup one of these would-be zero modes comes from the vortex core and the other from a domain wall mode circulating the vortex, as can be seen through an analysis similar to what we did for the edge modes. In terms of the quantum Hall physics, as we will see, the absence of the vortex Majorana modes indicates that these states have Abelian topological order.

  For our square lattice model, from counting the number of edge modes in open boundary conditions we found  $\mathcal{C}=2$ at $\kappa=nQ$ points in the quasi-continuum limit. We show in Fig.~\ref{fig:gapped dirac}(b) such a situation with $n=8$ with an open boundary  condition in $x$ direction. As can be seen, there are two propagating modes of each chirality.  In Appendix \ref{app:e-m} we compute the lattice corrections to $t, \tilde t$, and $t'$ in Eq.~\eqref{eq:ts} for our square lattice model, and show that indeed $\tilde t>t$ at $t'=0$. We have not done the calculation for other lattices, and from symmetry constraints alone both $\mathcal{C}=0$ and $\mathcal{C}=2$ phases are possible.
  Quite remarkably,   with a $p_x+ip_y$ local pairing symmetry, the PDW state realizes a band topology of that for a $d+id$ superconductor, even though their symmetry  properties are very different.

\subsubsection{gapped phase for $q>k_F$}

Now we consider the hybridization of bound states with $q>k_F=\sqrt{2m\mu}$ located at the nodes of the PDW order parameter. Here we show that the bulk spectrum of the 2D array is gapped, and it is topologically trivial.

For $q>k_F$, as we mentioned,  both \eqref{eq: ZM type 1} and \eqref{eq:ts} continue to hold. The only difference is that now $\kappa$, $N_o$, and $\sin(\kappa x)$ are imaginary. It is convenient to express \eqref{eq:ts} in terms of real variables:
\begin{align}
\label{eq:ts2}
t = -&\frac{\bar \kappa }{4m}\exp{(-q\lambda/2)}\nonumber\\&\times\left[2N_e\bar N_o\cosh\left (\frac{\bar\kappa\lambda}{2}\right)-(N_e^2+\bar N_o^2)\sinh\left(\frac{\bar\kappa\lambda}{2}\right)\right] \nonumber\\
\tilde{t} = -&\frac{\bar \kappa }{4m}\exp{(-q\lambda/2)}\nonumber\\&\times\left[2N_e\bar N_o\cosh\left (\frac{\bar\kappa\lambda}{2}\right)+(N_e^2+\bar N_o^2)\sinh\left(\frac{\bar\kappa\lambda}{2}\right)\right] \nonumber\\
\!\!\!t'= &\frac{\bar \kappa}{4m}\exp{(-q\lambda/2)}(N_e^2-\bar N_o^2)\sinh\left(\frac{\bar\kappa\lambda}{2}\right)
\end{align}
where $\bar \kappa \equiv |\kappa| = \sqrt{q^2-k_F^2}$, and 
\begin{align}
\label{eq:neno}
N_e = \sqrt{ \frac{2q(q^2 -\bar\kappa^2)}{2 q^2 - \bar \kappa^2}},~
\bar N_o = \sqrt{ \frac{2q(q^2 -\bar\kappa^2)}{\bar \kappa^2}},
\end{align}
With these, we notice that the Pfaffian of the spectrum \eqref{eq:fs}:
\begin{align}
&\mathrm{Pf}[\tilde{\mathcal{H}}(k_x,k_y)]\nonumber\\
&=(v_y k_y)^2 - 4(t'^2+t\tilde t) \cos^2(k_x/2)  + (t+\tilde t\,)^2 >0. \nonumber
\end{align}
\emph{always} hold for $q>k_F$. The proof is elementary: The Pfaffian is greater than $(t-\tilde t)^2-4t'^2 \propto \bar N_o^2 N_e^2>0$.  This indicates that the fermionic  spectrum is gapped for $q>k_F$. The size of the gap is of the same order as the $t$'s. We remind that the gap in $q>k_F$ regime is typically larger than the gap in $q<k_F$, since the latter is given by lattice corrections (see Appendix \ref{app:e-m}) and vanishes in the continuum limit.

The topology of this state can be obtained by a similar analysis at $q\gg k_F$. Since for smaller $q>k_F$ the gap does not close, the topology does not change. For $q\gg k_F$, $q\approx \bar \kappa$, and $N_e\approx \bar N_o$. Thus from \eqref{eq:ts2} we have $t'\to 0$. Then, as we discussed previously in Sec.~\ref{subsubsec: c=2} and shown in Fig.\ \ref{trivial phase}, the Chern number of this state again depends on the relative amplitude of $t$ and $\tilde t$. Here, from Eq.~\eqref{eq:ts2}, we have unambiguously $t>\tilde t$, and therefore the gapped state has $\mathcal{C}=0$, i.e., the band topology of the gapped state at $q>k_F$ is trivial.

It is worth  comparing the trivial pairing state we obtained with the ``strong pairing phase" considered in Ref.~\cite{readgreen2000}. As we cautioned, the ``strong pairing" there refers to a situation in which the ``normal state" does not have a FS ($\mu<0$). Both our state and the strong pairing phase are topologically trivial. In our case, however, we note that we have always set $\mu>0$ in our state, so it may seem puzzling at first why our state is trivial. Here the trivial topology is obtained by invoking additional domain wall states, which by themselves couple into a 2d system that neutralizes the total Chern number.

\section{Coexistence of PDW order and uniform pairing order}
\label{sec: Coexistence of stripe pairing order and uniform pairing order}
In this section we focus on the fermionic spectrum in the presence of coexisting PDW order parameter and uniform $p_x+ip_y$ pairing order parameter. We will refer to this state as the  $p_x+ip_y$ striped pairing state.

We determine the fermionic spectrum in the regime where the paired state has  a $p_x+ip_y$ PDW state coexisting with a uniform component of the $p_x+ip_y$ pairing order. In general, we find that the fermionic spectrum is gapped. In particular, for $Q<q<k_F$, the Majorana FS is gapped as the inversion symmetry is broken. We analyze the band topology of the gapped phases and present a phase diagram.

\begin{figure}
\includegraphics[width=\columnwidth]{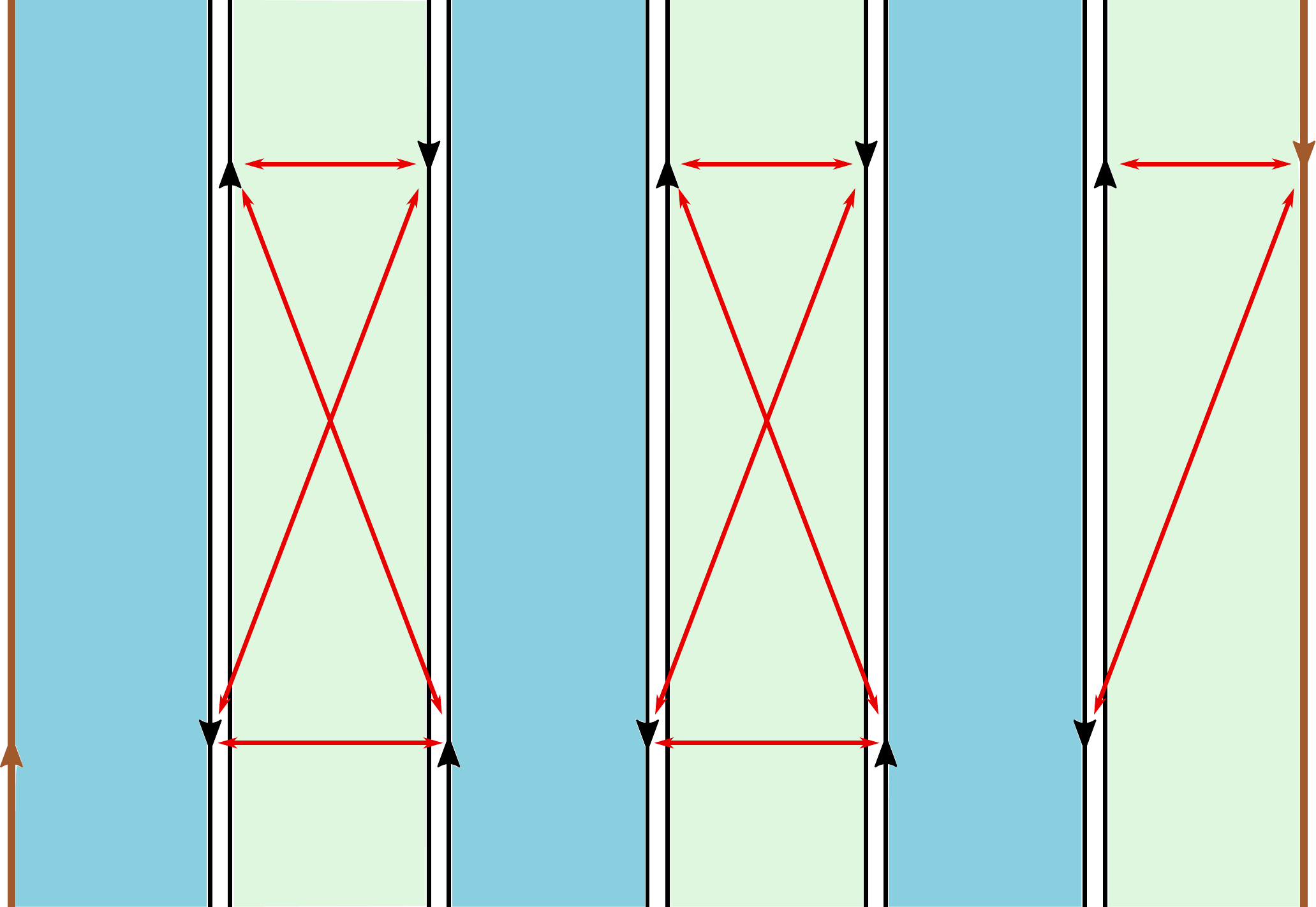}
\caption{Illustration of the gapped fermionic spectrum {in the coexistence phase of PDW and uniform pairing with a Chern number $\mathcal{C}=1$.}} 
\label{topological phase}
\end{figure}
\subsection{Gapping of the Majorana FS}
{We assume that the {order parameter of the} uniform component has the same phase as the overall phase for the order parameter of the PDW state}. 
The order parameter in real space is of the form
\begin{equation}
\label{eq: PDW order parameter}
\begin{split}
&\,
\Delta(x) 
=
\Delta_{\rm u}+\Delta_{\rm pdw} \Big[ 1 + \sum_{\epsilon = 1,2} \sum_{n \in \mathbb{Z}}  
(-1)^{\epsilon}\,\textrm{sgn}(x - x^{(\epsilon)}_{n}) \Big]
\,,
\end{split}
\end{equation}
{where $\Delta_{\rm u}$ represents a uniform component of the order parameter.}
Crucially, we see that the inversion symmetry centered at the DW's and ADW's with $x=x_n^{(\epsilon)}$ are now broken by the uniform component $\Delta_{\rm u}$. A direct consequence is that the Majorana FS for $q<k_F$ protected by the particle-hole symmetry $\mathcal{C}$ and inversion symmetry $\mathcal{I}$ (Sec.~\ref{sec:symmetry-protection-fs}), gets gapped. Indeed, numerical calculations on Eq.~\eqref{eq:lat1} with both $\Delta_{\rm u}$ and $\Delta_{\rm pdw}$ confirm that the fermionic spectrum is gapped.

Instead of a detailed evaluation of the hopping matrices in a tight-binding Hamiltonian, like we did for \eqref{eq: tight-binding H}, one can understand the gap opening in an intuitive way.
In Appendix~\ref{app:stability} we show that the two zero-mode solutions obtained in Sec.~\ref{sec:Domain-wall-bound-states} persist so long as $|\Delta_{\rm u}|<|\Delta_{\rm pdw}|$.
With $\Delta_{\rm u}$, the domains and anti-domains become ``imbalanced", with order parameters alternating between $\pm \Delta_{\rm pdw}+\Delta_{\rm u}$, and we assume $|\Delta_{\rm u}|<|\Delta_{\rm pdw}|$. As a direct result, the wave packets of the propagating modes bound on a DW at $x=0$ also becomes asymmetric. Following a similar procedure leading to Eq.~\eqref{eq: ZM type 1} and Eq.~\eqref{eq:LRmodes},  
\begin{align}
\label{eq:LRmodes3}
\langle{\vec r}\ket{\Psi'_R} \propto&\begin{cases} \exp[ik_y y-i\kappa_+ x] \exp[{-q_+x }], &\mathrm{ for~}x>0 \\ \exp[ik_y y-i\kappa_- x] \exp[{q_-x }], &\mathrm{ for~}x<0
\end{cases} \nonumber\\
\langle{\vec r}\ket{\Psi'_L} \propto &\begin{cases} \exp[ik_y y-i\kappa_+ x] \exp[{-q_+x }], &\mathrm{ for~}x>0 \\ \exp[ik_y y-i\kappa_- x] \exp[{q_-x }], &\mathrm{ for~}x<0
\end{cases}\end{align}
where
\begin{align}
q_\pm \equiv& m(\Delta_{\rm pdw} \pm \Delta_{\rm u}),\nonumber\\
\kappa_{\pm}  \equiv& \sqrt{k_F^2-q_{\pm}^2}.
\end{align}
Importantly, the wave packet of both left and right moving modes are more extended into the domain where the order parameter has a smaller magnitude. Indeed, this is expected since the local pairing order gaps out the local density of states and dictates the exponential decay of the wave  packet.

Intuitively, the coupling between domain wall states is stronger at regions with greater overlap of their wave functions. Analogous to the hopping amplitudes depicted in Fig.~\ref{fig:tibi}, one can define six hopping matrices $t_{\pm}$, $\tilde t_{\pm}$, and $t'_{\pm}$, where $\pm$ distinguishes domains with stronger or weaker local pairing order. Similar to Eq.~\eqref{eq:ts}, we have $t_{\pm}, \tilde{t}_{\pm}, t'_{\pm}\propto \exp{(-q_{\pm}\lambda/2})$. In the tight-binding limit, we then have $t_{-}\ll t_{+}$, $t'_{-}\ll t'_+$, and $\tilde{t}_-\ll \tilde{t}_+$. In this limit, the system is ``quadrumerized,'' with each quadrumer  being composed of the left and right moving modes at a DW-ADW pair. We illustrate this in Fig.\ \ref{topological phase}. The quadrumerization develops in regions with a smaller pairing order and hence greater overlap between wave packets. Each quadrumer consists of two left movers and two right movers, and the hybridization in their wave functions lead to a gap.

It turns out the such a coexistence state has nontrivial band topology {manifested by the presence of chiral edge states}. In the quadrumer (tight-binding) limit, we consider a finite system (see Fig.\ \ref{topological phase}). Depending on the termination of the finite system, near each physical edge there is either one unpaired chiral mode {(left edge in  Fig.~\ref{topological phase})} or three would-be chiral modes {(one edge mode and two nearby domain wall modes, right edge in Fig.~\ref{topological phase})} with a net chirality. Either way, one gapless chiral mode survives at each physical edge. The existence of the stable gapless modes near the edges indicates the this coexistence state is topological and has a Chern number $\mathcal{C}=1$. It belongs to the same universality class as the weak-coupling regime in Ref.\ \cite{readgreen2000}.

\subsection{Mean-field pairing phase diagram}}
\label{sec:global}

\begin{figure}[h]
\includegraphics[width=\columnwidth]{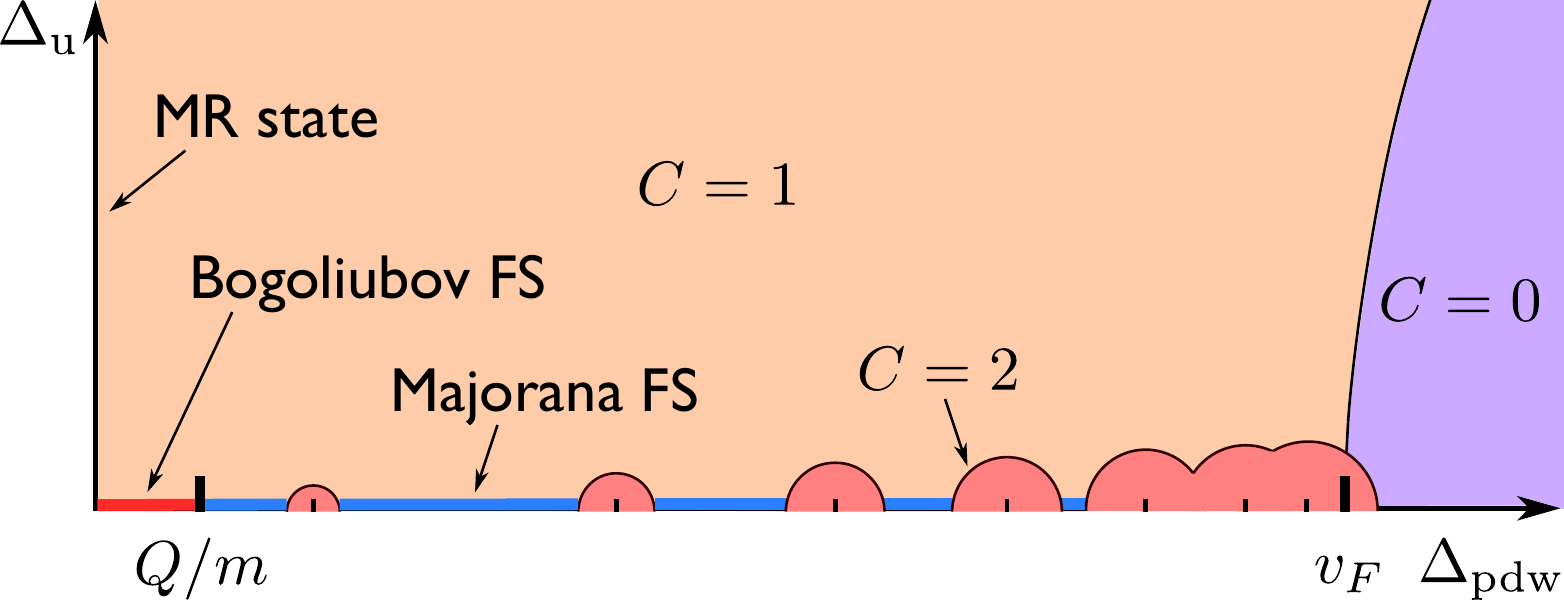}
\caption{Schematic pairing phase diagram for the fermionic states as a function of the PDW order parameter and a coexisting uniform $p$-wave order. When the uniform component $\Delta_\textrm{u} = 0$, the 
hybrididization of the bulk domain walls in general gives rise to a FS for $ \Delta_\textrm{pdw} <  v_F$. 
For $\Delta_{\rm pdw}<Q/m$ (or $q<Q$), the FS is from a perturbative reconstruction of the normal state FS. The fermionic excitations are Bogoliubov quasiparticles. For $Q/m<\Delta_{\rm pdw}<v_F$, the FS is made of Majorana modes from the domain walls. We use terms ``Bogoliubov FS'' and ``Majorana FS'' to distinguish them.
Near specific values of $\Delta_{\rm pdw}$ such that $\kappa = nQ$, (weak) lattice effects gaps out the fermionic spectrum with a Chern number $\mathcal{C}=2$ (although $\mathcal{C}=0$ states may also be possible depending on lattice details).
Above the critical pairing strength $\Delta_\textrm{pdw} > v_F$, the system enters a topologically trivial gapped state ($\mathcal{C}=0$). 
This state survives a finite amount of uniform component $\Delta_\textrm{u}$. The neutral FS becomes gapped for any $\Delta_\textrm{u} \neq 0$, when
the system enters the topological pairing phase ($\mathcal{C}=1$) whose edge states contain a chiral Majorana mode. For $\Delta_{\rm u} \gg \Delta_{\rm pdw}$ the system approaches a uniform $p$-wave state. }
\label{fig: phase-diagram}
\end{figure}

{We end this section by placing all the phases mentioned above in a phase diagram in terms of the PDW order parameter $\Delta_{\rm pdw}$ and a possible coexisting uniform $p_x+ip_y$-wave order parameter $\Delta_{\rm u}$. We summarize the results in Fig.~\ref{fig: phase-diagram}.}

We have carefully analyzed the gapped phases in a pure PDW state, both for $q<k_F$ (or equivalently $\Delta_{\rm pdw} <v_F$) and for $q>k_F$ (or equivalently $\Delta_{\rm pdw} >v_F$). Due to the spectral gap, these states are stable in the presence of a small $\Delta_{\rm u}$, which induces a ``competing mass" that leads to a $\mathcal{C}=1$ phase. One naturally expects that the sizes of these phases in $\Delta_{\rm u}$ direction is proportional to their spectral gaps. Therefore, in a semi-continuous limit where the lattice corrections are small, the $\mathcal{C}=2$ phase with $q<k_F$ occupies a much smaller region with $\Delta_{\rm u}\neq 0$ than  the $\mathcal{C}=0$ phase with $q>k_F$ does. Both of these phase transitions involves a change in Chern number by 1, and we have numerically verified that the phase transition occurs with a gap closing through a Dirac point at the phase boundaries.

\section{The $p_x+ip_y$ PDW fractional quantum Hall states}
\label{sec:PDW-FQH}

Our study so far addressed the properties of the fermion spectrum in a paired state and, as such,
can be viewed as a description of a striped superconductor with chiral $p$-wave order parameter. 
We now turn on the implications of our results for the FQH physics of this state,  keeping in mind that a paired FQH state
is not a superconductor, but in fact a charge insulator in an applied magnetic field. 
In order to make contact with the physics of the paired quantum Hall states, we reintroduce both charge and neutral modes on equal
footing, and recall that they are coupled to a dynamical Chern-Simons gauge field. 

The neutral fermion modes we studied in the previous sections, which originated from a change in
sign of the $p_x+ip_y$ order parameter, are akin to zero energy Andreev bound states in a Josephson junction,
where the difference in the phase of the order parameter is $\pi$. Just as an external magnetic flux alters the phase difference and
gives rise to a spatially oscillating current passing through a Josephson junction,~\cite{tinkham-book} one might worry that
the same would happen in this case due to the Chern-Simons and the external magnetic fields.
The situation, however, is greatly simplified (at least in the mean field description assumed here) due to the 
complete screening of the external magnetic flux by the Chern-Simons flux attached to the particles, which implies
that the total effective magnetic field experienced by the composite fermions 
is zero and, thus, the gauge fields do not alter the character of the Andreev bound states. 

\begin{figure}
\includegraphics[width=.6\columnwidth]{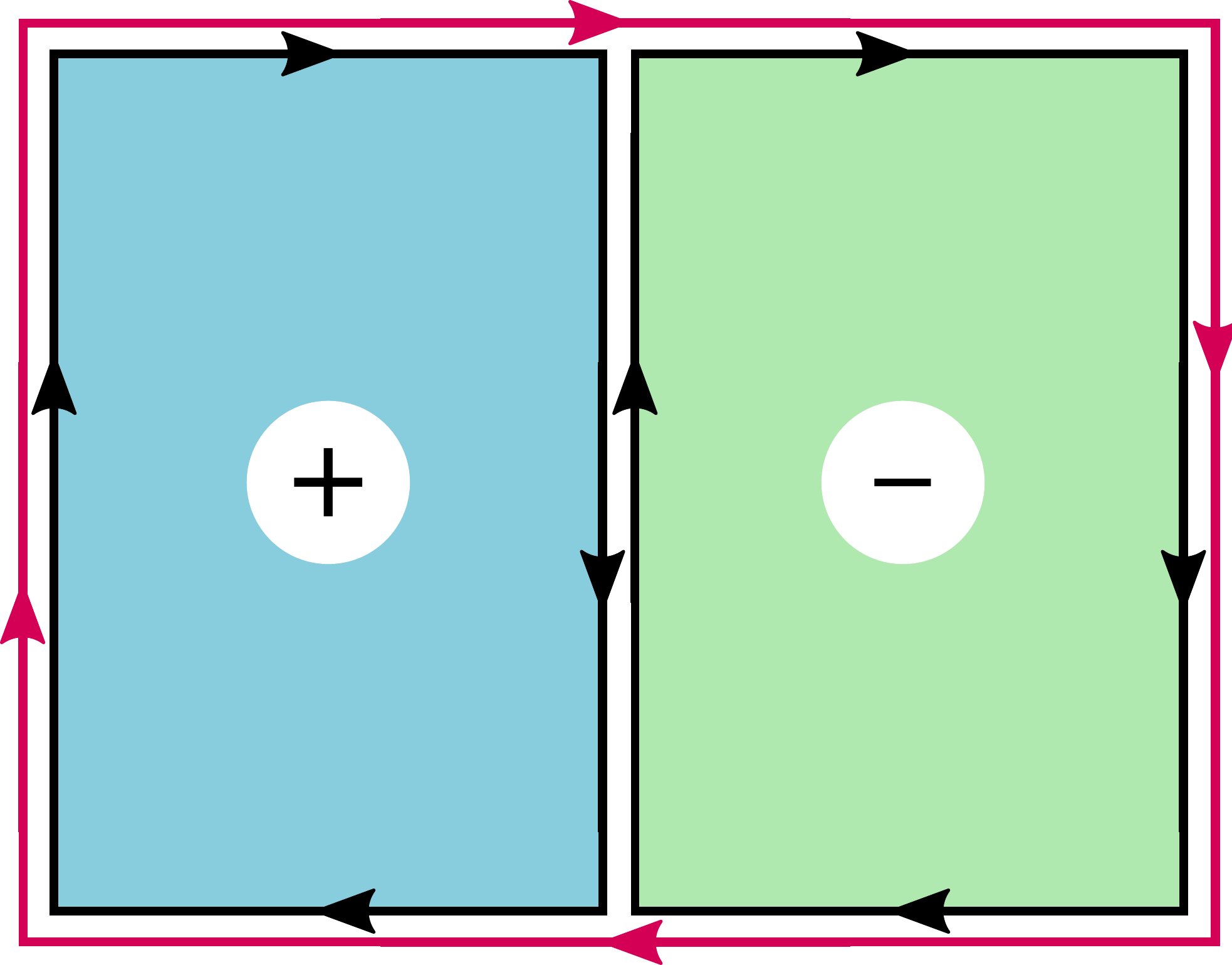}
\caption{Charge chiral mode at the boundary and Majorana modes both at the edge and at the domain wall in the bulk of the system.}
\label{fig: single bulk dw}
\end{figure}

The discussion above can be made more concrete by recalling that 
at filling fraction $\nu=1/2$ of this $N=1$ LL, upon performing a standard mapping to composite fermions coupled 
to a fluctuating Chern-Simons gauge fields $a_{\mu}$, with $\mu = 0,1,2$, the effective Lagrangian of the paired FQH state 
(in units where $e=\hbar=c=1$)reads
\begin{align}
\label{eq:paired-FQH-lagrangian}
\mathcal{L} &\,
= 
{\psi^{\dagger}\,( i D_{0} + \mu )\, \psi}
+
\frac{1}{2m} |\vec{D}\psi|^{2}
+
\Delta\,\psi \psi + \Delta^{*}\,\psi^{\dagger} \psi^{\dagger}
\nonumber\\
&\, 
+ \frac{1}{4\pi}\frac{1}{2}\,\epsilon^{\mu\nu\lambda} a_{\mu} \partial_{\nu} a_{\lambda}
\end{align}
{where, 
$D_{\mu} = \partial_\mu + i(A_{\mu} + a_{\mu})$}
and, on average,
\begin{equation}
\label{eq:flux-cancellation}
\vec{\nabla} \times (\vec{a} + \vec{A}) = 0.
\end{equation}
This condition defines the mean field state and enforces that the electronic density
$\rho = 1/2$ everywhere in the bulk of the system.
Had the total flux $\vec{\nabla} \times (\vec{a} + \vec{A})$
been non-zero in the region across the domain wall (which would have implied a local variation either of the magnetic field, the charge density, or both), 
then the associated Josephson effect would have depended on the gauge invariant phase difference across the junction (i.e., the domain wall)
that carries a contribution from the gauge fields. However, in the mean field state characterized by Eq.~\eqref{eq:flux-cancellation}
the phase difference $\pi$ associate with the order parameter $\Delta$ fully specifies the properties
of the low energy states bound at the domain walls. To simplify notation, in the Lagrangian of Eq.\eqref{eq:paired-FQH-lagrangian} the $p_x+ip_y$ symmetry structure of the pairing has been included in the pair field $\Delta$.

\subsection{Spectra of  $p_x+ip_y$ PDW FQH states}

The bulk Chern-Simons term in Eq.~\eqref{eq:paired-FQH-lagrangian} encodes the property that the system is a charge insulator in bulk
with a gapless chiral bosonic mode at the boundary of the system describing the charged excitations. The neutral fermion excitations
of the system, either in the bulk or at the boundary, on the other hand, are described by the fermionic sector with the PDW order parameter. 
Thus, the neutral fermionic spectrum of the striped paired FQH states are those that we obtained for the $p_x+ip_y$ PDW state 
in Sec.~\ref{sec:Properties-Fermionic}, while the charged bosonic sector is described by the Chern-Simons action. The striped paired FQH system
then has gapless neutral excitations supported at domain walls in the bulk of the system, while remaining a bulk charge insulator with gapless charge modes on the edge,
as illustrated in Fig.~\ref{fig: single bulk dw} (showing only two domains).

The analysis of Sec.~\ref{sec:Properties-Fermionic} combined with the charge sector discussed above, 
shows that there are four phases of the striped paired FQH state,
which are summarized by Fig.~\ref{fig: phases} and the pairing phase diagram Fig.~\ref{fig: phase-diagram}. 
In the absence of a uniform $p_x+ip_y$-wave component $\Delta_{\rm u}$,
when the Fermi energy is large compared to the pairing term of the PDW order ($k_F > q$, or equivalently $v_F>\Delta_{\rm pdw}$)
and the system supports domain walls in the bulk,
then the zero modes in each domain wall hybridize with their neighbors giving rise to a 2D FS of charge-neutral Bogoliubov quasiparticles, a Majorana FS, as represented in Fig.~\ref{fig: phases}(c).
Quite remarkably, these neutral Majorana excitations are formed while the charged degrees of freedom remain gapped (which implies that tunneling of electrons in the bulk is suppressed by the charge gap). This neutral FS implies the system has an anisotropic unquantized bulk thermal conductivity, and a heat capacity that scales linearly with temperature $T$, while its charge transport is gapped with a sharp plateau of $\sigma_{xy}=1/2$. This exotic ``critical phase'' is one of our central findings of this work. 

{A different paired stripe FQH state at $\nu = 5/2$ was proposed by Wan and Yang \cite{Wan-2016}, which is a state with alternating domains of Pfaffian and anti-Pfaffian states. Similar to our results, they  found a state with gapped charge modes but gapless neutral modes at each domain wall. However, the domain wall between the Pfaffian and the anti-Pfaffian state has a more intricate structure that in the case of the $p_x+ip_y$ PDW state we prose here, leading to a more complex set of domain-wall modes.  Moreover, the analysis of Ref.~\cite{Wan-2016}  neglects the coupling (and tunneling) between the neighboring domain wall modes which, as we showed here, plays an important role in the physics of the state. Thus, it is an open question whether these couplings will induce a bulk gap or not. In contrast, the gapless state obtained here \emph{survives} the coupling between the domain wall modes, as is protected by symmetry.}

Furthermore, as indicated in the phase diagram Fig.~\ref{fig: phase-diagram}, for $k_F > q$, centered around each would be Dirac points at $\sqrt{k^{2}_F - q^2} = n Q$, $n\in \mathbb{Z}$, there exists a gapped phase with Chern number $\mathcal{C}=2$ with two co-propagating neutral models near the boundary (in addition to the charge mode). This $\mathcal{C}=2$ region  represents an Abelian FQH state, as the vortices do not support Majorana zero modes. The edge CFT is composed of two chiral Majorana fermions and one charge mode, with a chiral central charge $c=2$. We are not aware of any previous discussions of this exotic FQH state.

The neutral FS is unstable towards gapped phases with distinct topological properties. 
The first type of instability happens in the weak pairing regime ($k_F > q$) and it is triggered by a non-zero
uniform component of the $p_x+ip_y$-wave order parameter $\Delta_{\rm u} \neq 0$. In this case, the neutral FS becomes topologically gapped with a Chern number $\mathcal{C}=1$,
and the system is in the same universality class as the non-Abelian Pfaffian state. The transition between this state and the aforementioned $\mathcal{C}=2$ state is of Dirac type. Just like the Pfaffian state, in the bulk there exists non-Abelian anyons with $e/4$ electric charge and the edge is described by a $U(1)_2\times \mathrm{Ising}/\mathbb{Z}_2$ CFT with a chiral central charge $c=3/2$. {The factor of $\mathbb{Z}_2$ accounts gauge symmetry associated with representing the electron operator
as a product of a Majorana fermion of the Ising sector and a charge one vertex operator of the $U(1)_2$ sector.}

Another instability of the neutral FS occurs at $\Delta_{\rm u} = 0$ 
when the pairing potential is stronger than the Fermi energy, $q > k_F$. This transition is associated with a qualitative change in the character
of the DW zero mode states, as discussed in Sec.~\ref{sec:Domain-wall-bound-states}, which causes right and left moving modes to display an asymmetric decay near the domains and, consequently, gaps both the bulk and the edge modes. This pairing phase is characterized by a Chern number $\mathcal{C}=0$.
The disappearance of the neutral fermion modes from the low energy spectrum indicates a transition from a non-Abelian state to an Abelian state~\cite{Kane-Stern-Halperin-2017}, the latter in the universality class of the Halperin paired state~\cite{halperin1983}, where electrons form tightly bound charge $2$ pairs that condense in an Abelian state with $\sigma_{xy} = 1/2$. 

Note that, unlike in Ref.~\cite{readgreen2000}, where the transition from the Pfaffian to the Abelian state only occurs at chemical potential $\mu=0$ for a spatially uniform order parameter, for the PDW state considered here, the critical phase occurs for a positive $\mu$, {and for a \emph{finite} range of the parameter $\Delta_{\rm pdw}$.} The strength of PDW order parameter behaves as a new ``knob'' that tunes the system through that
transition between different topological orders. As we explained, this striking stability of the neutral FS stems from the symmetries possessed by the PDW state, which restricts the coupling of the Majorana modes both within each domain wall and between domain walls.

\subsection{Phase structure near $\nu=5/2$}
\label{sec:phase-diagram}

We end with a qualitative discussion of the place of the $p_x+ip_y$-PDW FQH state in a  global phase diagram of quantum Hall states. Much as in the case of other liquid-crystalline quantum Hall states \cite{FK}, the  $p_x+ip_y$-PDW FQH state can melt either quantum mechanically or thermally in a number of different ways, similar to the melting phase diagram conjectured for the PDW superconductor in Ref.~\cite{Berg-2009b}, by a generalization of the well-known theories of 2D classical melting \cite{Kosterlitz-1973,Nelson-1979,Young-1979}.

In the case of the PDW superconductor (including a $p_x+ip_y$-PDW  state), the different pathways are also determined by the proliferation of the panoply of its topological excitations.  The $p_x+ip_y$ PDW, just as its $d$-wave cousin, has three types of topological excitations: quantized vortices, half-vortices bound to single dislocations, and double-dislocations~\cite{Berg-2009b}. The proliferation of quantized and/or half-vortices destroy the paired state and lead to two possible compressible unquantized states: either a charge stripe state or a compressible nematic phase.

{On the other hand, the proliferation of double dislocations leads to an uniform incompressible state best described as a {\em quartet FQH condensate}. The quartet FQH condensate is  an analog of the charge-$4e$ superconductor~\cite{Berg-2009b}, where four (rather than two) fermions form a bound state and condense. Strong arguments have been presented \cite{CJWang-2016} that a quartet condensate (as well as a charge-$4e$ topological superconductor) has Abelian topological order. A detailed analysis of the properties of the quartet FQH state, however, is beyond the scope of the present work.}

However, the properties of the different resulting phases depend on features specific to the physics of the FQH states. In addition to the condensates (paired or not), FQH fluids have a dynamical emergent gauge field, the Chern-Simons gauge field. One of the consequences of the emergent {\em Chern-Simons} gauge field being dynamical is that the vortices of the condensate (i.e. the fundamental quasiparticles of the FQH state) have finite energy, instead of the logarithmically divergent energy of a vortex of a neutral superfluid. On the other hand the effective interaction between the vortices may be attractive (as in a type I superconductor) or repulsive (as in a type II superconductor). In addition, FQH vortices carry fractional charge and, hence, vortices also interact with each other through the Coulomb interaction. The interplay between these different interactions was analyzed in the context of uniform paired FQH states by Parameswaran and coworkers \cite{Parameswaran-2011,Parameswaran-2012a}, who predicted a complex phase with different liquid-crystal phases depending on {whether the FQH fluid is in a type I or type II regime.}

{Much of the analysis summarized above} can be extended, with some caveats,  to the case of the $p_x+ip_y$-PDW FQH state. One important difference vis-\'a-vis the PDW superconductors is that in a 2D system such as the 2DEG, in the absence of an underlying lattice the dislocations of the associated charge order cost a finite amount of free energy. As such they proliferate at any finite temperature, thus restoring translation invariance, and resulting in a nematic phase at all non-zero temperatures \cite{Toner-1980}. This problem was considered before in the context of high temperature superconductors in Ref.~\cite{Barci-2011}. However, in the presence of {strong enough anisotropy, e.g. by uniaxial strain or by a tilted magnetic field, can trigger a phase transition to a state  with unidirectional order which can be a $p_x+ip_y$ PDW FQH state or a charge stripe state (the latter case was found in a DMRG numerical work of Zhu and coworkers \cite{Zhu-2017}). Both of these stripe states thermally melt by proliferating dislocations, whose interactions are logarithmic in an anisotropic system~\cite{Toner-1980}. }

The precise interplay between these (and other) phases depends on details of the length scales that govern quantum Hall fluids. It is widely believed (for  good reasons!) that in the lowest Landau level all length scales are approximately of the same order of magnitude as the magnetic length. In Landau levels $N\geq 1$ and higher, other scales may come into play. This fact is evinced by the recent experiments near $\nu=5/2$ which find an interplay between a (presumably uniform) paired state and a compressible nematic phase \cite{samkharadze2016}, and between a compressible nematic phase and a stripe phase (albeit in the $N=2$ Landau level)  \cite{qian2017}.

These additional length scales may affect the structure of the vortices and of the other topological excitations, and therefore the nature of the state obtained for fields and/or densities away from the precise value of the filling fraction $\nu=5/2$, but still inside the plateau for the incompressible state. More specifically,  the FQH state has a fluctuating gauge field, with  a Chern-Simons term and a (subdominant) Maxwell term, which introduces a screening-length in the problem which will affect the structure of the vortices, ``type I'' or ``type II''. This problem was considered before in the context of  relativistic field theory \cite{Paul-1986}, and, more relevant to our analysis, in the context of paired FQH states {\cite{Parameswaran-2011,Parameswaran-2012a},  
although they did not consider the interplay of a possible $p_x+ip_y$ paired state. For example, if a ``type II'' regime may become accessible, the vortex states may exhibit intertwined orders of analogous to  those that arise in high $T_c$ superconductors \cite{Edkins-2018,Wang-2018b}.  In this case, a $p_x+ip_y$ PDW phase may arise in the vortex ``halos'' of the uniform paired state, and could be stabilized close to  $\nu=5/2$.
The upshot of this analysis is that a complex phase diagram may yet to be uncovered, beyond what has been seen in  recent experiments.

\section{Discussion and conclusion}
\label{sec:conclusions}

In this paper we have studied the properties of a 2D pair-density wave  state
with a $p_x+ip_y$ chiral order parameter, which is periodically varying along one direction, and have shown that 
this physical system can support exotic bulk symmetry-protected (gapless or gapped) fermionic spectrum. This bulk gapless phase results from
the hybridization of pairs of counter-propagating Majorana fermion states localized near the nodes of the order parameter. The stability
of the Majorana  states near the domain walls is a consequence of a combination of inversion and chiral symmetries associated with 
the unidirectional PDW order parameter. 

In the weak coupling regime (in the BCS sense) characterized by $v_F > \Delta_{\textrm{pdw}}$, the zero modes are localized within the distance $q^{-1}$, where 
$q = m\,\Delta_{\textrm{pdw}}$. We have shown that the hybridization of these domain wall modes gives rise to a Majorana FS that is protected
by both particle-hole and inversion symmetries and that the robustness of the FS can be captured by the properties of a Pfaffian. 
Our findings have been supported both by an effective theory valid in the regime $q \gg Q$, 
in which the low energy modes on adjacent domain walls hybridize weakly, as well as by numerical calculations in the regime where
the domain walls strongly couple to many neighboring domain walls. 

The FS obtained in the $v_F>\Delta_{\rm pdw}$ regime is generically unstable to the presence of perturbations that break inversion symmetry.
In particular, a small uniform component of the order parameter breaks the inversion symmetry that maps $\Delta_{\textrm{pdw}} \rightarrow -\Delta_{\textrm{pdw}}$
around a domain wall and destroys the FS, giving rise to a gapped spectrum of neutral fermionic excitations. Moreover, we have shown that 
this gapped phase is topological as it supports a chiral Majorana branch at the boundary of the system, which has the same topological properties
of the uniform $p_x+ip_y$ paired state. Our analysis has also shown the existence of special points characterized by the condition $\sqrt{k^{2}_{F}-q^2} = nQ$ ($n \in \mathbb{Z}$), for which FS becomes a Dirac point at $(k_x, k_y) = (0,0)$. This Dirac point is a consequence of the continuum approximation of the band structure
and generically becomes gapped by distortions of the Majorana wavefunctions due to lattice effects, where the system, interestingly, has a fermionic spectrum with Chern number $\mathcal{C}=2$, and thus supports two edge Majorana modes. On the other hand, for the strong coupling limit $\Delta_{\rm pdw}>v_F$, we found the resulting fermionic spectrum to be trivial. These findings have been summarized in the phase diagram Fig.~\ref{fig: phase-diagram}.

Viewed as a striped superconductor, our theory shows the existence of zero energy extended Majorana states in the bulk
of a PDW phase with chiral $p$-wave order parameter. In this case, all the excitations of the systems are neutral Majorana modes. 
We applied this theory to the
 paired FQH state at filling $\nu=5/2$ in which the composite fermions pair into a state with a spatially dependent order parameter.
In fact, recent numerical work~\cite{lee2018} has shown that, as a function of the 2DEG layer thickness, the effective interactions experienced by composite fermions
in $N \geq 1$ Landau levels can give rise to a Pomeranchuk instability, which could account for a mechanism behind the formation of a nematic FQH state as it is, in fact, in line with recent experimental findings.~\cite{samkharadze2016}

In our description of the striped FQH state at $\nu=5/2$, the charge modes remain gapped in the bulk and give rise to a chiral bosonic density mode at the 
boundary, which is a conformal field theory with central charge $c=1$. The PDW order parameter changes only the properties of the neutral fermionic sector.

From the discussion above, in the weak coupling regime, the neutral particles develop and gapless FS protected by symmetry while the bulk
remains gapped to charge excitations. Consequently, while the tunneling of neutral (Majorana) quasiparticles is facilitated by the absence of an energy gap in the bulk,
the tunneling of electron is suppressed by the charge gap. Moreover, a non-zero uniform component gaps the neutral fermionic spectrum and the system develops
a chiral Majorana branch; we then identify this phase as a striped Moore-Read state. 
At $\sqrt{k_F^2-q^2}=nQ$ points, the edge CFT includes two Majorana branches, and the topological order becomes Abelian.
On the other hand, when the pairing effects become sufficiently strong,
the system becomes gapped (even in the absence of a uniform component) and the systems enters a phase without a neutral Majorana edge state; this phase is then identified
with the striped (Abelian) Halperin paired state. 

We close with a discussion of the possible relation between the $p_x+ip_y$ PDW FQH state and the very recent experiments of Hossain and coworkers \cite{Hossain-2018}, whose results were posted on the Physics Archive after this work was completed. This experiment considers a 2DEG in an AlAs heterostructure which has two elliptical electron pockets oriented at $90^\circ$ degrees of each other. Each pocket has very anisotropic effective masses, with a ratio of 5:1. Under a very weak unidirectional strain field, the Landau level of one or the other pocket is emptied and the system has a strong electronic anisotropy. Importantly, in these systems, at the fields in which the experiments are done, the Zeeman energy is larger than the Landau gap, as also is the energy splitting due to the applied strain.

{Remarkably, the experiments of Ref.~\cite{Hossain-2018} find a clear plateau in the $N=1$ Landau level at $\nu=3/2$, equivalent of the much studied $\nu=5/2$ plateau in the 2DEG in GaAs-AlAs heterostructures. However, these authors also found a remarkable transport anisotropy {\em inside the plateau regime}, by which, below some well-defined temperature, the longitudinal resistance $R_{xx}$ (along the $(100)$ direction) rises sharply to a value comparable to $R_{xy}$, while resistance $R_{yy}$ (along the $(0,1,0)$ direction) decreases sharply. This nematic behavior is reminiscent to the earlier findings of Xia and coworkers \cite{xia2011} near filling fraction $\nu=7/3$  in the $N=1$ Landau Level of the 2DEG in GaAs-AlAs heterostructures.}

{ While it is tempting to interpret these experimental results as evidence for the existence of the $p_x+ip_y$ PDW FQH state, it also raises a puzzle since the magnitude of the longitudinal resistance seems incompatible with this state which has a bulk charge gap. We should note that this experiment cannot distinguish a nematic state (which is uniform) from any stripe state (which breaks translational symmetry), paired or not.  There are several possible ways to understand this behavior. One is that for a sample with the form of a QH bar the strain does not  force the system into a single oriented domain but that there may be two orthogonally  oriented domains in the bar geometry. In this scenario, the longitudinal transport is only carried by the charge edge mode and it is drastically anisotropic. Other  scenarios are also possible, such as the one suggested by the analysis of Parameswaran and coworkers \cite{Parameswaran-2012a}, perhaps the paired state is in the ``type I'' regime which leads to a form of Coulomb frustrated phase separation. However, in this latter scenario, it is hard to understand why $R_{xy}$ has a sharp plateau. At any rate, if the state found in these experiments is a $p_x+ip_y$ PDW FQH state it should exhibit bulk thermal conduction, as predicted by our analysis.}

In summary, we have presented a new scenario characterized by 
a 2D chiral topological phase being intertwined with a striped
order, in which low energy neutral fermionic degrees of freedom are found to be supported at the 
nodes of the PDW order parameter. Our findings have implications both for the understanding of 
nematic paired FQH states at filling $\nu=5/2$, as well as for 
nematic (or striped) superconductors.

Note: After this work was completed we became aware of a preprint by Barkman and coworkers \cite{barkman2018} who considered a time-reversal invariant $p$-wave superconductor consisting of alternating domains with $p_x\pm i p_y$ pairing. The physics of this  state is very different of the time-reversal breaking $p_x+ip_y$ PDW superconductor that we present in this paper.

\acknowledgments{We thank Daniel Agterberg, Steven Kivelson, Ganpathy Murthy, Mansour Shayegan, and Ajit Srivastava for discussions. {EF is particularly grateful to S. Kivelson for numerous discussions (and the suggestion for the interpretation of the anisotropic transport in the context of the $7/3$ state.)} This work was supported in part by the Gordon and Betty Moore Foundation EPiQS Initiative through Grant No. GBMF4305 at the University of Illinois (LHS and YW) and the National Science Foundation grant No. DMR-1725401 at the University of Illinois (EF). LHS and YW performed part of this work at the Aspen Center for Physics, which is supported by National Science Foundation grant PHY-1607611.}

\onecolumngrid

\appendix

\section{Majorana fermions at the nodes of the pair-density wave state}
\label{app:nodes}

We provide here details of the Majorana fermion states located at the nodes of the PDW order parameter. 
For that we begin with the BdG Hamiltonian of the paired state
\begin{equation}
\label{eq: BdG H - appendix}
H = 
\begin{pmatrix}
\epsilon(\vec{p})-\mu  & \frac{1}{2} \{  p_{-}, \Delta \}
\\
\frac{1}{2} \{  p_{+}, \Delta^{*} \} & -\epsilon(\vec{p})+\mu
\end{pmatrix}	
\,,
\end{equation}
where $\epsilon(\vec{p}) = \vec{p}^{2}/2m$,
$\vec{p} = (p_x,p_y) = (-i \partial_x , -i \partial_y)$ and $p_{\pm} = p_x \pm i p_y = -i\partial_{\pm}$ (We set $\hbar = 1$). 
The anticommutators appearing in the BdG Hamiltonian can be expressed as
$
\frac{1}{2} \{  p_{-}, \Delta \} = -\frac{i}{2} \partial_{-} \Delta - i\Delta \partial_{-}
$
and
$
\frac{1}{2} \{  p_{+}, \Delta^{*} \} = -\frac{i}{2} \partial_{+} \Delta^{*} - i\Delta^{*} \partial_{+}
$.
The system is defined on the plane with $x \in (-\infty,\infty)$ and $y \in (-L/2,L/2)$.
The BdG Hamiltonian \eqref{eq: BdG H - appendix} possesses a particle-hole symmetry (redundancy)
$
\sigma_1 H \sigma_1 = - H^{*}	
$,
which relates positive and negative energy states: if $\Psi_E$ is an eigenmode
of $H$ with energy $E$, then $\sigma_1 \Psi^{*}_{E}$ is an eigemode with energy $-E$.
For an order parameter $\Delta(x) \in \mathbb{R}$, Hamiltonian \eqref{eq: BdG H - appendix} is translation invariant along the y direction, 
such that the momentum eigenmodes $\Psi_{E,k_{y}}(x,y) = e^{i k_{y} y} \phi_{E,k_{y}}(x)$ satisfy
\begin{equation}
\label{eq: BdG H - appendix v3}
\begin{split}
&\,
H_{k_{y}} \phi_{E,k_{y}}(x) = E \, \phi_{E,k_{y}}(x)
\,,
\\
&\,
H_{k_{y}}
=
\begin{pmatrix}
\frac{-\partial^{2}_{x}+k_{y}^{2}}{2m}-\mu  & -\frac{i}{2} \Delta'(x) + \Delta(x)(-i \partial_x - i k_{y})
\\
-\frac{i}{2} \Delta'(x) + \Delta(x)(-i \partial_x + i k_{y}) & \frac{\partial^{2}_{x} - k_{y}^{2}}{2m} + \mu
\end{pmatrix}	
\,.
\end{split}
\end{equation}

\subsection{Zero Modes}

We now consider an order parameter that changes sign at $x = 0$ and we look for Majorana fermions
supported along this domain wall. Particle-hole symmetry of \eqref{eq: BdG H - appendix} implies that Majorana modes are described by
\begin{equation}	
\label{eq: ZM chirality - appendix}
\Psi_0^{\epsilon} = 
\begin{pmatrix}
\phi_\epsilon
\\
(-1)^{\epsilon+1} \, \phi^{*}_{\epsilon}
\end{pmatrix}
\,,
\quad
\epsilon = 1,2
\,,
\end{equation}
(where $\epsilon$ denotes the zero modes chirality) and,
according to \eqref{eq: ZM chirality - appendix}, the zero mode equation, after setting $k_y=0$ in \eqref{eq: BdG H - appendix v3}, simplifies to
\begin{equation}
\label{eq: ZM eq v1 - appendix}
\left( - \frac{\partial^{2}_{x}}{2m} - \mu  \right) \phi_{1,2}(x) \mp i \Delta(x) \partial_{x} \phi^{*}_{1,2}(x) 
\mp \frac{i}{2} \Delta'(x) \phi^{*}_{1,2}(x) = 0	
\,.
\end{equation}
We now consider the sharp domain wall configuration
\begin{equation}
\label{eq: domain wall - appendix}
\Delta_{\pm}(x) = \mp \Delta_{\textrm{pdw}} \,\textrm{sgn}(x)	
\,,
\Delta'_{\pm}(x) = \mp 2 \Delta_{\textrm{pdw}} \, \delta(x)
\,,
\Delta_{\textrm{pdw}} > 0
\,,
\end{equation}
and solve for the zero mode with positive/negative chiralities (the $\pm$ label in the order parameter is intended to show that it is correlated with the zero mode chirality). With the ansatz 
\begin{equation}
\Psi_{\epsilon}(x)  = e^{i\pi/4} \, \frac{\psi(x)}{\sqrt{2}}\,
\begin{pmatrix}
1
\\
(-1)^{\epsilon+1} i
\end{pmatrix}
\equiv
e^{i\pi/4}\,
\psi(x)\,u_{\epsilon}
\,,
\end{equation}
where $u_{\epsilon}$ are the eigenspinors of $\sigma_y$,
the zero mode equation \eqref{eq: ZM eq v1 - appendix} reads
\begin{equation}
\label{eq: ZM eq v2}
\left( - \frac{\partial^{2}_{x}}{2m} - \mu  \right) \psi(x) -\Delta_{\textrm{pdw}} \textrm{sgn}(x) \partial_{x} \psi(x) 
-\Delta_{\textrm{pdw}} \delta(x) \psi(x) = 0	
\,.
\end{equation}

We first consider $x<0$ and plug the ansatz $\psi_{L} \sim e^{i P_L x}$ into \eqref{eq: ZM eq v2}, leading to
$
\frac{P^{2}_{L}}{2m} - \mu + i \Delta_{\textrm{pdw}} \, P_{L}= 0	
$,
whose solutions are
\begin{equation}
\begin{split}
&\,
P_{L} = -i (q \pm \bar\kappa)
\,,
\quad
q = m \Delta_{\textrm{pdw}} > 0
\,, 
\quad
\bar\kappa = \sqrt{q^2-2m\mu} > 0
\,.
\end{split}
\end{equation}
Then, as long as $\mu>0$, the solution is normalizable in the $x<0$ half space,
and the solution for $x<0$ reads ($\bar\kappa\equiv i\kappa$)
\begin{equation}
\label{eq: ZM negative x}
\psi_{L}(x) = e^{q x} \left[ A \cos{(\kappa x)} + B \sin{(\kappa x)}   \right]
\,,
\quad
x < 0
\,.
\end{equation}

Similar approach to the $x>0$ region leads to
\begin{equation}
\label{eq: ZM positive x}
\psi_{R}(x) = e^{-q x} \left[ A \cos{(\kappa x)} + C \sin{(\kappa x)}   \right]
\,,
\quad
x > 0
\,.
\end{equation}
Notice that continuity of $\psi(x)$ at $x=0$ fixes the same coefficient $A$ in \eqref{eq: ZM negative x} and \eqref{eq: ZM positive x} and that the asymptotic form of $\psi_{L/R}$ for large values of $|x|$ guarantees that the solution is normalizable.
The boundary condition at $x=0$ is delt with by integrating \eqref{eq: ZM eq v2} in an infinitesimal interval $( -\epsilon, + \epsilon )$ around $x=0$ and invoking continuity of $\psi(x)$ at $x=0$, which yields
\begin{equation}
\label{eq: ZM condition at origin}
-\frac{1}{2m} \textrm{lim}_{\epsilon \rightarrow 0 } \left[ \psi'(+\epsilon) - \psi'(-\epsilon) \right]
-\Delta_{\rm pdw}\,\psi(0)
=
0
\,.	
\end{equation}

\subsection{Majorana fermions for $q<k_F$}

For $k_F^2\equiv 2m\mu > q^2$, there are \textit{two} orthonormal zero energy solutions for a given $\epsilon = 1,2$.
\begin{subequations}
\label{eq: ZM type 1 - appendix}
\begin{equation}
\label{eq: even ZM cosine}
\begin{split}
&\,
\Psi_{\epsilon,e}(x) = N_{e}\,\frac{e^{-q|x|}}{\sqrt{L}}\,\cos{(\kappa\,x)}\,u_{\epsilon}
\,,
\quad
N_{e} = \sqrt{\frac{2q(\kappa^2 + q^2)}{\kappa^2 + 2\,q^2} }
\,,
\end{split}	
\end{equation}
and
\begin{equation}
\label{eq: odd ZM sine}
\begin{split}
&\,
\Psi_{\epsilon,o}(x) = N_{o}\,\frac{e^{-q|x|}}{\sqrt{L}}\,\sin{(\kappa\,x)}\,u_{\epsilon}
\,,
\quad
N_{o} = \sqrt{  \frac{2q(\kappa^2 + q^2)}{\kappa^2} }
\,.
	\end{split}	
\end{equation}
\end{subequations}

\subsection{Majorana fermions for $q>k_F$}

For $q>k_F$, The expressions in \eqref{eq: even ZM cosine} and \eqref{eq: odd ZM sine} are still correct, but it is convenient to re-express them in real parameters:
\begin{subequations}
\label{eq: ZM type 2 - appendix}
\begin{equation}
\label{eq: even ZM cosh}
\begin{split}
&\,
\Psi_{\epsilon,e}(x) = \bar N_{e}\,\frac{e^{-q|x|}}{\sqrt{L}}\,\cosh{(\bar{\kappa}\,x)}\,u_{\epsilon}
\,,
\quad
\bar N_{e} = \sqrt{\frac{2q(q^2 -\bar{\kappa}^2)}{2\,q^2 -\bar{\kappa}^2} }
\,,
\end{split}	
\end{equation}
and
\begin{equation}
\label{eq: odd ZM sinh}
\begin{split}
&\,
\Psi_{\epsilon,o}(x) = \bar N_{o}\,\frac{e^{-q|x|}}{\sqrt{L}}\,\sinh{(\bar{\kappa}\,x)}\,u_{\epsilon}
\,,
\quad
\bar N_{o} = \sqrt{  \frac{2q(q^2 -\bar{\kappa}^2)}{\bar{\kappa}^2} }
\,.
\end{split}	
\end{equation}
\end{subequations}

\section{Stability of the Zero modes in the presence of a small uniform $p$-wave component}
\label{app:stability}

In the presence of a uniform component $\Delta_{\textrm{u}}$ of the order parameter, the domain wall is described by
\begin{equation}
\Delta(x) = \Delta_{\textrm{u}} -\Delta_{\textrm{pdw}}\,\textrm{sgn}(x)
\,,
\end{equation}
with $\Delta_{\textrm{u}} > 0$ and $\Delta_{\textrm{pdw}} > 0$ . We now show that the zero energy solutions are stable as long as 
$\Delta_{\textrm{u}} < \Delta_{\textrm{pdw}}$. 
To see that, we note that for $x<0$, the order parameter is $\Delta_{\textrm{pdw}} + \Delta_{\textrm{u}} \equiv \Delta_{L}$ 
and for $x>0$, we have $-\Delta_{\textrm{pdw}} + \Delta_{\textrm{U}} \equiv -\Delta_{R}$, where $\Delta_{L/R} > 0$. Defining 
$
q_{L} = m\,\Delta_{L}
$,
$
q_{R} = m\,\Delta_{R}
$,
$
\kappa_{L} = \sqrt{2\,m\,\mu - q^{2}_{L}} 
$
and
$
\kappa_{R} = \sqrt{2\,m\,\mu - q^{2}_{R}}
$,
the zero mode solutions have the form
\begin{equation}
\begin{split}
&\,
x < 0: \quad
\psi(x) = e^{q_{L}\,x}
\,
\left[
A\,\cos{(\kappa_{L}\,x)}
+
B\,\sin{(\kappa_{L}\,x)}
\right]/\sqrt{L}
\\
&\,
x > 0: \quad
\psi(x) = e^{-q_{R}\,x}
\,
\left[
A\,\cos{(\kappa_{R}\,x)}
+
C\,\sin{(\kappa_{R}\,x)}
\right]/\sqrt{L}
\,
\end{split}
\end{equation}
and satisfy the condition
\begin{equation}
-\frac{1}{2m}\,
\textrm{lim}_{\epsilon \rightarrow 0}\,
\Big[
\psi'(+\epsilon)
-
\psi'(-\epsilon)
\Big]
-\Delta_{\rm pdw}\,\psi(0) = 0
\,,
\end{equation}
which implies 
\begin{equation}
\kappa_{R}\,C = \kappa_{L}\,B
\,.
\end{equation}

We then identify two \textit{orthogonal} solutions $\psi_{1}(x)$ and $\psi_{2}(x)$ given by
\begin{equation}
\psi_{1}(x)
=
A\,
\Big[
\Theta(-x)\,e^{q_{L}\,x}\,\cos{(\kappa_{L}\,x)}
+
\Theta(x)\,e^{-q_{R}\,x}\,\cos{(\kappa_{R}\,x)}
\Big]
\,,
\end{equation}
\begin{equation}
\psi_{2}(x)
=
B\,
\Big[
\Theta(-x)\,e^{q_{L}\,x}\,\sin{(\kappa_{L}\,x)}
+
\frac{\kappa_{L}}{\kappa_{R}}\,
\Theta(x)\,e^{-q_{R}\,x}\,\sin{(\kappa_{R}\,x)}
\Big]
\,,
\end{equation}
where $A$ and $B$ are normalization constants that can be readily determined.

Notice that, in the limit $\Delta_{\textrm{u}} \rightarrow 0$,
the solutions above reduce to the even and odd parity solutions obtained before.

\section{Derivation of the tight-binding effective Hamiltonian Eq.~\eqref{eq: tight-binding H}}
\label{app:tight-binding}

Let us consider a tight-binding variational state
\begin{equation}
\label{eq: Bloch state - appendix}
\ket{\Psi_{k_{x},k_{y}}}
=
\sum_{\epsilon = 1,2} \sum_{\mu = L,R}
c_{\epsilon,\mu}
\sum_{n \in \mathbb{Z}}
\frac{e^{i k_{x}\,n\,\lambda}}{\sqrt{N}} 
\ket{\Psi_{\epsilon,\mu,n}(k_y)}
\,,
\end{equation}
where the subscript $\epsilon =1$ denotes DW modes and $\epsilon=2$ ADW modes. 
Recall that $\ket{\Psi_{\epsilon=1}}\propto \ket{u_1} = (1,i)^T/\sqrt{2}$ and $\ket{\Psi_{\epsilon=2}}\propto \ket{u_2} =(1,-i)^T/\sqrt{2}$.
The coefficients $c_{\epsilon,\mu}$ are variational parameters, which are determined by minimizing
the energy $E_{\vec{k}} = \bra{\Psi_{k_{x},k_{y}}} H \ket{\Psi_{k_{x},k_{y}}} / \bra{\Psi_{k_{x},k_{y}}} \Psi_{k_{x},k_{y}} \rangle$,
which yields the secular equation
\begin{equation}
\label{eq: secular equation - appendix}
\mathrm{det}
\Big[
\mathcal{O}^{-1}(\vec{k})\,\mathcal{H}(\vec{k}) - E_{\vec{k}}\,I_{4 \times 4}
\Big] = 0
\,,
\end{equation}
where
\begin{equation}
\mathcal{O}(\vec{k})_{\varepsilon_1 \mu_1, \varepsilon_2 \mu_2}
=
\sum_{n} e^{i k_x n} \langle \Psi_{\varepsilon_1 \mu_1, -n}(k_y) | \Psi_{\varepsilon_2 \mu_2, 0}(k_y) \rangle
\end{equation}
\begin{equation}
\mathcal{H}(\vec{k})_{\varepsilon_1 \mu_1, \varepsilon_2 \mu_2}
=
\sum_{n} e^{i k_x n} \langle \Psi_{\varepsilon_1 \mu_1, -n}(k_y) | 
E_{\mu_2}(k_y) + V_{\rm total}(x) - v^{(\varepsilon_2)}_{0}(x) |
\Psi_{\varepsilon_2 \mu_2, 0}(k_y) \rangle
\end{equation}

\begin{subequations}
\begin{equation}
\begin{split}
&\, 
V_{\rm total}(x) = \frac{1}{2}\,\{ -i\,\partial_{x} , \Delta_{\rm total}(x) \}\,\sigma_{x}
\,,
\quad
v^{(\varepsilon_2)}_{0}(x) = \frac{1}{2}\,\{ -i\,\partial_{x} , \Delta^{(\varepsilon_2)}_{0}(x) \}\,\sigma_{x}
\end{split}
\,,
\end{equation}
\begin{equation}
\begin{split}
\Delta_{\rm total}(x) &\, =
\Delta_{\rm pdw}
+
\sum_{n}\,\Delta^{(1)}_n(x)
+
\sum_{n}\,\Delta^{(2)}_n(x)
\\
&\,=
\Delta_{\rm pdw} - \sum_{n}\,\Delta_{\rm pdw}\,\textrm{sgn}(x - n\,\lambda) + \sum_{n}\,\Delta_{\rm pdw}\,\textrm{sgn}(x - (n+1/2)\,\lambda)  
\,.
\end{split}
\end{equation}
\end{subequations}

We first notice that, because DWs and ADWs are respectively proportional to the orthonormal spinors $\ket{u_1}$ and $\ket{u_2}$, 
the overlap matrix $\mathcal{O}$ is the identity matrix to leading order:
\begin{equation}
\mathcal{O}
=
I_{4 \times 4}
+
\textrm{O}(e^{-q\,\lambda})
\,.
\end{equation}
Then, according to Eq.~\eqref{eq: secular equation - appendix}, the band structure is obtained by diagonalizing the Hermitian matrix
\begin{equation}
\label{eq: TB effective H - PDW}
\mathcal{H}(k_x,k_y)
=
\begin{pmatrix}
\mathcal{H}_{1 R\, , \, 1 R}
&
\mathcal{H}_{1 R\, , \, 1 L}
&
\mathcal{H}_{1 R\, , \, 2 R}
&
\mathcal{H}_{1 R\, , \, 2 L}
\\
\mathcal{H}_{1 L\, , \, 1 R}
&
\mathcal{H}_{1 L\, , \, 1 L}
&
\mathcal{H}_{1 L\, , \, 2 R}
&
\mathcal{H}_{1 L\, , \, 2 L}
\\
\mathcal{H}_{2 R\, , \, 1 R}
&
\mathcal{H}_{2 R\, , \, 1 L}
&
\mathcal{H}_{2 R\, , \, 2 R}
&
\mathcal{H}_{2 R\, , \, 2 L}
\\
\mathcal{H}_{2 L\, , \, 1 R}
&
\mathcal{H}_{2 L\, , \, 1 L}
&
\mathcal{H}_{2 L\, , \, 2 R}
&
\mathcal{H}_{2 L\, , \, 2 L}
\,.
\end{pmatrix}
\end{equation}

\subsection{Diagonal Matrix Elements of \eqref{eq: TB effective H - PDW}}

We have for $\mathcal{H}_{1 R\, , \, 1 R}$
\begin{equation}
\begin{split}
\mathcal{H}_{1 R\, , \, 1 R}(k_x,k_y)
&\,=
\sum_{n}\,e^{i k_x \,\lambda\,n}
\,
\int_{x}\,
\Psi^{*}_{1 R}(x+n\,\lambda)\,\,\bra{u_{1}}\,
\Big[
E_{R}(k_y)
+
V_{\rm total}(x) - v^{(1)}(x)
\Big]
\ket{u_{1}}
\,
\Psi_{1 R}(x)
\\
&\,=
E_{R}(k_y)
\sum_{n}\,e^{i k_x\,\lambda\,n}
\,
\int_{x}\,
\Psi^{*}_{1 R}(x+n\,\lambda)\,
\phi_{1 R}(x)
\\
&\,=
E_{R}(k_y)
\Big[
1 + \textrm{O}(e^{-q\,\lambda})
\Big]
\approx E_{R}(k_y)
\,,
\end{split}
\end{equation}
where, in passing from the first to the second line, we used 
$
\bra{u_{1}}\,
\Big[
V_{\rm total}(x) - v^{(1)}(x)
\Big]
\ket{u_{1}}
=
0
$
due to $V_{\rm total}(x) - v^{(1)}(x)$ being proportional to the Pauli matrix $\sigma_x$. 
Similar calculation leads to the expression for the diagonal matrix elements of Eq.~\eqref{eq: TB effective H - PDW}:
\begin{equation}
\mathcal{H}_{1 R/L \, , \, 1 R/L} = \pm E_{R}(k_y) 
\,,
\quad
\mathcal{H}_{2 R/L \, , \, 2 R/L} = \pm E_{R}(k_y)
\,.
\end{equation}

\subsection{Off-diagonal Matrix Elements of \eqref{eq: TB effective H - PDW}}

We can show, by the same reasoning as before, that, to leading order, the following matrix elements are zero:
\begin{equation}
\mathcal{H}_{1 R/L \, , \, 1 L/R} = 0
\,,\quad
\mathcal{H}_{2 R/L \, , \, 2 L/R} = 0
\,.
\end{equation}

We are then left with the non-zero off-diagonal matrix elements of \eqref{eq: TB effective H - PDW},
$\mathcal{H}_{2 \mu_{2}\, , \, 1 \mu_{1}}(k)$, with $\mu_{1}, \mu_{2} = R/L$. To leading order
\begin{equation}
\begin{split}
\mathcal{H}_{2 \mu_{2}\, , \, 1 \mu_{1}}(k_x,k_y)
&\,=
R_{\mu_{2} \,,\, \mu_{1}}\,e^{i k_x\,\lambda}
+
S_{\mu_{2} \,,\, \mu_{1}}
+
\textrm{O}(e^{-q\,\lambda})
\,,
\end{split}
\end{equation}
where
\begin{subequations}
\label{eq: def R and L - app}
\begin{equation}
R_{\mu_{2} \,,\, \mu_{1}}
=
-\Delta_{\textrm{pdw}} \int ^{-\lambda/2}_{-\infty}dx
\left[
\Psi^{*}_{2 \mu_2}(x+\lambda/2) \partial_{x} \Psi_{1 \mu_1}(x)
-
\partial_{x} \Psi^{*}_{2 \mu_2}(x+\lambda/2) \Psi_{1 \mu_1}(x)
\right]
\,.
\end{equation}
\begin{equation}
S_{\mu_{2} \,,\, \mu_{1}}
=
\Delta_{\textrm{pdw}} \int^{\infty}_{\lambda/2} dx
\left[
\Psi^{*}_{2 \mu_2}(x-\lambda/2) \partial_{x} \Psi_{1 \mu_1}(x)
-
\partial_{x} \Psi^{*}_{2 \mu_2}(x-\lambda/2) \Psi_{1 \mu_1}(x)
\,.
\right]
\end{equation}
\end{subequations}
Evaluation of the integrals Eq.~\ref{eq: def R and L - app} gives the effective Hamiltonian
\begin{equation}
\mathcal{H}(k_x,k_y)
=
\begin{pmatrix}
v_{y}k_y
& 
0 
& 
t' + t'e^{-ik_x} 
&
t- \tilde{t}e^{-ik_x} 
\\
0 
&
-v_{y}k_y
&
-\tilde{t} + te^{-ik_x} 
&
t' + t'e^{-ik_x} 
\\
t' + t'e^{ik_x} 
&
-\tilde{t} + te^{ik_x} 
&
v_{y}k_y
&
0
\\
t - \tilde{t}e^{ik_x} 
&
t' + t'e^{ik_x} 
&
0
&
-v_{y}k_y
\end{pmatrix}
,
\label{eq: tight-binding H - app}
\end{equation}
where the parameters $t$, $\tilde{t}$ and $t'$ are given by Eq.~\ref{eq:ts}.

\section{Lattice corrections to the hopping matrices $t$, $\tilde t$, and $t'$}
\label{app:e-m}
In this appendix we compute the leading order corrections to Eq.~\eqref{eq:ts} by an underlying square lattice. We will focus on the quasi-continuous limit, where the Fermi wavelength $\lambda_{F}\equiv 2\pi/k_F$ is much larger than the lattice constant $a$.

The wave function of the domain wall modes can be obtained by solving the lattice version of \eqref{eq: ZM eq v1}, and by using a exponential function ansatz, the even- and odd-parity wave functions still satisfy Eq.~\eqref{eq: ZM type 1}, only the expression for $q$, $\kappa$, and $N_{o,e}$ are different from their continuum version. By a simple analysis these lattice corrections are of $O[(k_Fa)^2 ]$.

We recall that the hopping amplitudes, for example $t'$ was obtained by an integral
\begin{align}
t'= -\Delta_{\rm pdw}\int_{\lambda/2}^{\infty}dx \Big(&\left[\partial_x\Psi_{L}^*(\vec r)\right] \Psi_{R}(\vec r-\lambda/2)-  \Psi_{L}^*(\vec r) \left[\partial_x\Psi_{R}(\vec r-\lambda/2)\right]\Big)
\end{align}
For a lattice system, first one needs to replace $\partial_x$ with its lattice version $i\sin(\hat k_x)$, and doing so introduces corrections of $O[(k_Fa)^2 ]$. Besides, one should replace the integral with summations at the lattice sites. The leading correction from this replacement can be obtained from the Euler-Maclaurin formula
\begin{align}
\int_{\lambda/2}^{\infty} f(x) dx =& a\left[f\left(\frac{\lambda}{2}+\frac{a}{2}\right)+f\left(\frac{\lambda}{2}+\frac{3a}{2}\right)+\cdots\right] +a^2f'\left(\frac{\lambda}{2}+\frac a 2\right) + O(a^3).
\end{align}
Then including the leading-order Euler-Maclaurin correction, $t'$ is found to be
\begin{align}
t'=&-\frac{\kappa }{4m}\exp\left(-\frac{q\lambda}{2}\right)\left(N_e^2+N_o^2\right )\sin\left(\frac{\kappa\lambda}{2}\right)  \nonumber\\
&-\frac{q^2}{2m}\exp\left(-\frac{q\lambda}{2}\right)\left[N_e^2(2\delta(0) + q)+2N_eN_o\kappa\right]a^2.
\end{align}
Regularizing $\delta(0) = 1/a$, we see that the leading correction to $t'$ is of  $O(k_Fa)$ ($q\lesssim k_F$), given by the $\delta$-function term. We do not need to keep all other $O[(k_Fa)^2]$ terms. 
Including the lattice corrections for all  couplings we have
\begin{align}
t = -&\frac{\kappa }{4m}\exp\left(-\frac{q\lambda}{2}\right)\left[2N_eN_o\cos\left (\frac{\kappa\lambda}{2}\right)+(N_e^2-N_o^2)\sin\left(\frac{\kappa\lambda}{2}\right)\right]\nonumber\\& - \frac{q^2a}{m}\exp\left(-\frac{q\lambda}{2}\right)N_e^2  \nonumber\\
\tilde{t} = -&\frac{\kappa }{4m}\exp\left(-\frac{q\lambda}{2}\right) \left[2N_eN_o\cos\left (\frac{\kappa\lambda}{2}\right)-(N_e^2-N_o^2)\sin\left(\frac{\kappa\lambda}{2}\right)\right]\nonumber\\& - \frac{q^2a}{m}\exp\left(-\frac{q\lambda}{2}\right)N_e^2  \nonumber\\
\!\!\!t'= -&\frac{\kappa }{4m}\exp\left(-\frac{q\lambda}{2}\right)(N_e^2+N_o^2)\sin\left(\frac{\kappa\lambda}{2}\right)\nonumber\\& - \frac{q^2a}{m}\exp\left(-\frac{q\lambda}{2}\right)N_e^2 .
\end{align}

In the main text we are interested in the case where $t'=0$. It is straightforward to  verify that in this case $\sin(\kappa\lambda/2)<0$, and $\tilde t>t$. From the criterion given in the main text, the Chern number of this phase is $\mathcal{C}=2$.

 \twocolumngrid

\bibliography{p+ip-stripes.bib}

\end{document}